\documentclass[12pt,draftclsnofoot,onecolumn]{IEEEtran}

\usepackage{amsfonts,amsthm,amssymb}
\usepackage[cmex10]{amsmath}
\usepackage[usenames,dvipsnames]{pstricks}
\usepackage[dvips]{graphicx}
\usepackage{subfigure}
\usepackage{mathtools}

\usepackage{xcolor}
\usepackage{stfloats}
\usepackage{empheq}
\newtheorem{lemma}{Lemma}

\newcommand{\txs}[1]{\textsc{#1}}
\newcommand{\txm}[1]{\textrm{#1}}

\newcommand{\Ip}{\mathcal{I}_p}
\newcommand{\Is}{\mathcal{I}_s}
\newcommand{\Q}{\mathcal{Q}}
\newcommand{\Pp}{\mathcal{P}_p}
\newcommand{\Ps}{\mathcal{P}_s}

\newcommand{\brc}[1]{\left( #1 \right)}
\newcommand{\sqbrc}[1]{\left[ #1 \right]}
\newcommand{\figbrc}[1]{\left\{ #1 \right\} }

\IEEEoverridecommandlockouts

\begin{document}

\title{Cooperative Communication Using \\ Network Coding
}
\author{
\IEEEauthorblockN{Nan Li, Lars K. Rasmussen and Ming Xiao\\}
\thanks{The authors are with the School of Electrical Engineering and ACCESS Linnaeus Center, KTH Royal Institute of Technology, Stockholm, Sweden, Email:\{nanli2, lkra, mingx\}@kth.se}
}
\maketitle

\begin{abstract}
We consider a cognitive radio network scenario where a primary transmitter and a secondary transmitter, respectively, communicate a message to their respective primary receiver and secondary receiver over a packet-based wireless link, using a joint automatic-repeat-request (ARQ) error control scheme. The secondary transmitter assists in the retransmission of the primary message, which improves the primary performance, and is granted limited access to the transmission resources. Conventional ARQ, as well as two network-coding schemes are investigated for application in the retransmission phase; namely the static network-coding (SNC) scheme and the adaptive network-coding (ANC) scheme. For each scheme we analyze the transmission process by investigating the distribution of the number of transmission attempts and approximate it by normal distributions. Considering both the cases of an adaptive frame size and a truncated frame size, we derive analytical results on packet throughput and infer that the ANC scheme outperforms the SNC scheme.
\end{abstract}



\section{Introduction}\label{sec:intro}
Cognitive radio \cite{Mitola} has received considerable attention as a potential means  to mitigate the growing pressure on limited attractive spectrum resources. Within the cognitive-radio paradigm knowledge of spectrum usage can be intelligently collected and utilised to improve spectrum utilisation \cite{xG}. In cognitive radio networks, multiple transmitter/receiver pairs from so-called primary and secondary co-existing systems may cooperate to obtain communal benefits. Consequently, the combination of user cooperation and cognitive capabilities for improving both spectrum utilisation and transmission performance has been considered. Cooperative relaying \cite{relay}, in particular, has been comprehensively considered. In this case the relay node is required to have some level of information about the source message being transmitted in order to successfully forwarding it to the destination. A cognitive node may be able of acquiring such information from the source node or by listening to the channel. In some works dedicated relay nodes are part of the network and are typically equipped with cognitive abilities. An example of such cognition is the relay node in \cite{A_relay},  which is able to decode both primary and secondary signals. Using an opportunistic adaptive relaying scheme the relay can decide whom to cooperate with, the primary or secondary transmission, or simultaneously assist both. With multiple relays, as in \cite{B_relays}, the best relay is selected by an adaptive cooperation diversity scheme to improve the performance of secondary transmissions, while ensuring the \emph{Quality of Service} (QoS) of the primary communication. In other works, the secondary system accesses the spectrum along with the primary system and cooperates to transmit as a relay. Both \emph{Amplify-and-Forward} (AF) \cite{AF} and \emph{Decode-and-Forward} (DF) \cite{ST_DF} relaying are studied to facilitate secondary usage of spectrum.

In this paper we focus on delay-insensitive data network services, where \emph{Automatic Repeat reQuest} (ARQ) schemes are typically applied for packet error recovery. With error-control coding and feedback, ARQ enables the application of network coding in broadcast \cite{BC} and multicast \cite{RelayMC} networks. In this context, network coding has a strong potential to improve network throughput, efficiency and scalability. Here intermediate nodes combine several packets for transmission, instead of simply relaying the packets of information they receive. Furthermore, Birk and Kol proposed network coding in multiple-unicast networks \cite{BK98,BK06} for efficiently supplying different data packets from a central server to multiple caching clients. We advance this view by establishing a cognitive radio network with cooperative transmission by the secondary transmitter over a broadcast channel. We explore in particular that the secondary transmitter is able to receive and decode the primary message, as well as combining its own message with the primary message in a network-coded transmission to increase the efficiency of both systems. In that context, the primary system can be assisted by allowing the secondary system accessed to limited spectral resources. In other words, if the secondary system assists in maintaining, or even improving, the primary system performance, a share of the bandwidth will be granted for its own transmission. 

In our previous work \cite{LXR_VTC14}, two network-coding schemes were investigated for use in the retransmission phase: namely, the static network coding scheme (SNC) and the adaptive network-coding scheme (ANC). The respective performances were favorably compared to a plain ARQ scheme. In the SNC scheme, the packet combining process is predetermined, which is suboptimal, while in the ANC scheme the combining process is adapted to the instantaneous acknowledgments received. In this paper, we further analyze the advantages of the ANC scheme by providing a lower bound on the throughput performance and comparing to the SNC scheme. Moreover, we investigate the performance of each scheme for two cases based on different constraints on the instantaneous frame size. In each case, we analyze three transmission sessions of the transmission process and approximate the distribution of the number of transmission attempts by a normal distribution to reduce the computational complexity.

Unless otherwise defined, the following notational rules are used. Random variables are uppercase boldface italic ($\boldsymbol{B}$), realisations of random variables and constants are uppercase italic ($B$), and sets are uppercase calligraphic ($\Q_p$). The probability mass function (pmf) of the random variable $\boldsymbol{B}$ is denoted $P_{\boldsymbol{B}}(B)$, the probability of the event $B> \widehat{B}$ is denoted as $\mathbb{P}\{B > \widehat{B}\}$, and the expectation of a random variable $B$ is denoted as $\overline{B} = E_{\boldsymbol{B}}[B]$. The negative binomial distribution, with parameters $B$ (total number of trials), $N$ (number of successes) and $p$ (probability of failure), is denoted as $\mathcal{NB}(B,N,p)$, and provides the distribution of the total number of independent and identically distributed Bernoulli trials before a specified deterministic number of successes occurs. The pmf is $P_{\boldsymbol{B}}(B)=\binom{B-1}{N-1}p^{B-N}(1-p)^N$ for fixed $N$ and $p$, and the mean value is $\overline{B}=N/(1-p)$. The normal distribution with mean $\mu$ and variance $\sigma^2$ is denoted as $\mathcal{N}\brc{\mu, \sigma^2}$, based on which, the truncated normal distribution with an upper limit $\widehat{B}$ is denoted as $\mathcal{TN}\brc{\hat{\mu},\hat{\sigma};\widehat{B}}$, where mean $\hat{\mu}$ and variance $\hat{\sigma}^2$ can be derived by $\mu$ and $\sigma^2$. Variables related to the analysis of conventional ARQ are distinguished by a superscript C, the SNC scheme by a superscript S, and the ANC scheme by a superscript A.

This paper is organised as follows. The cognitive radio network model is defined in Section \ref{sec:sysmodel}, and the two cooperation-based network coding schemes for multiple unicast transmissions are defined. A thorough performance analysis in terms of throughput and outage probability is detailed in Section \ref{sec:performance_a} and \ref{sec:performance_p}, where the transmission process is analyzed subject to the effects of different assumptions on the frame size. 
Numerical results are provided in Section \ref{sec:numresults}, and conclusions are given in Section \ref{sec:conclusion}.

\section{System Model}\label{sec:sysmodel}


\begin{figure}[tp]
\centering
\includegraphics[width=0.7\linewidth]{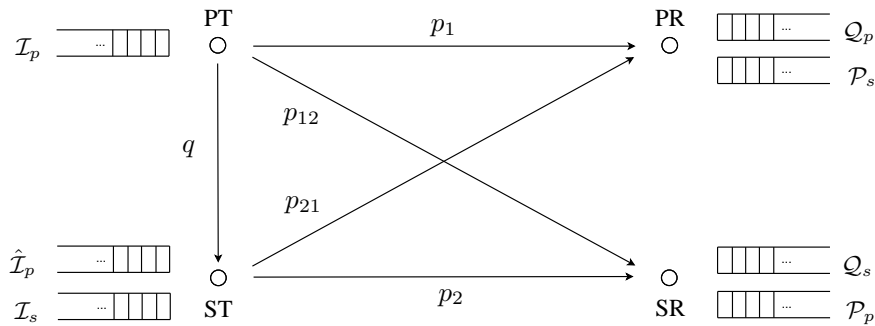}
\caption{Cognitive radio system with packet caches.}
\label{fig1:sys}
\end{figure}

We consider a cognitive radio network consisting of a single \emph{secondary transmitter} (ST) and \emph{secondary receiver} (SR) pair coexisting with a single \emph{primary transmitter} (PT) and \emph{primary receiver} (PR) pair, as shown in Fig. \ref{fig1:sys}. Each transmitter has information to be delivered to the given receiver. Here, the ST cooperates as a relay to assist in delivering the \emph{primary message} ($\Ip$), while in return accessing a share of the licensed resources to transmit the \emph{secondary message} ($\Is$).
All the links in the network are modelled as packet erasure links (PEL) with constant packet loss probabilities. Here, $p_1$ and $p_2$ in Fig. \ref{fig1:sys} denote the packet loss probabilities for the direct links between respective transmitters/receivers, while $p_{12}$ and $p_{21}$ denote the packet loss probabilities for the cross links, and $q$ denotes the packet loss probability for the link between the PT and the ST. A static channel model is considered where all the packet loss probabilities are assumed constant during the transmission process and known to the transmitters. We assume that each transmitter is aware of all packet losses in the network through ARQ acknowledgements (Ack/Nack), where all Ack/Nacks are instantaneous and error-free for simplicity. $\Q_p$ is the set of successfully received packets from $\Ip$ at the PR, where $\overline{\Q}_p$ is the complement of $\Q_p$, thus denoting the lost packets at the PR; the SR and the ST also receive packets from $\Ip$ which are stored in $\Pp$ and $\widehat{\mathcal{I}}_p$, respectively. Similarly for $\Is$, $\Q_s$ is the set of successfully received packets at the SR and $\Ps$ for the received packets at the PR. The primary message comprises of $N_p$ packets, and the secondary message of $N_s$ packets, denoted as $\Ip = \{\txm{I}_p^i \mid i=1,2, ... ,N_p\}$ and $\Is = \{\txm{I}_s^i \mid i=1,2, ... ,N_s\}$.
The notation is summarised in Table \ref{table1}.
\begin{table}[h]
\renewcommand{\arraystretch}{1.5}
\caption{Notation}
\label{table1}
\centering
\begin{tabular}{c|c|c|c}
\hline
\bfseries Notation & \bfseries Description & \bfseries Notation & \bfseries Description\\
\hline\hline
$\Ip$ & Primary message & $\Q_p$ & Primary packets received at PR\\
\hline
$\Is$ & Secondary message & $\Q_s$ & Secondary packets received at SR\\
\hline
$\widehat{\mathcal{I}}_p$ & Primary packets received at ST & $\Pp$ & Primary packets received at SR\\
\hline
& & $\Ps$ & secondary packets received at PR \\
\hline
\end{tabular}
\end{table}

\begin{figure}[htb]
\centering
\includegraphics[width=0.6\linewidth]
{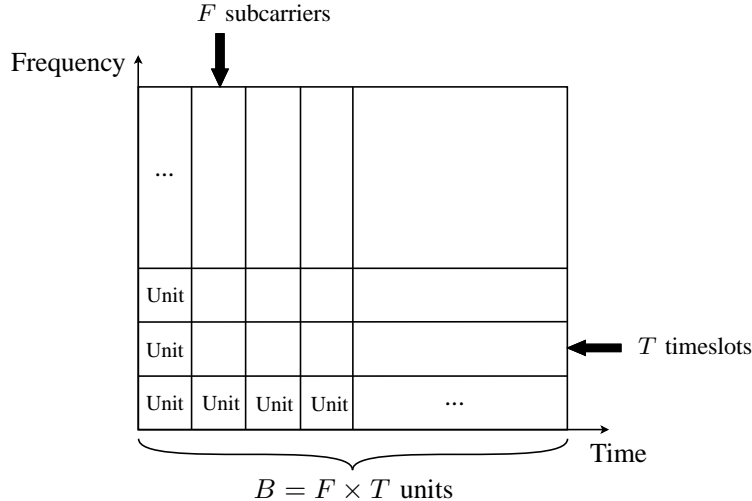}
\caption{Frame structure.}
\label{fig2:frame}
\end{figure}

For ease of exposition, we model the available spectrum resources in terms of identical \emph{resource units} (RUs), representing a time-frequency block, where all packets are of equal size and each packet can be transmitted within one resource unit. A general time-frequency frame model for an \emph{Orthogonal Frequency-Division Multiple Access} (OFDMA) system is shown in Fig. \ref{fig2:frame}. By properly adjusting the allocation of subcarriers, transmit power and constellation sizes \cite{OFDM_packet}, a packet can be transmitted within one resource unit. Compared to \emph{Time Division Multiple Access} (TDMA) and \emph{Frequency-Division Multiple Access} (FDMA), OFDMA provides better flexibility for scheduling the resources and lower delay as compared to TDMA.

Consider a frame of size $B$ RUs that are shared between the primary and secondary systems through three transmission sessions. Note that the size of each session is constrained to be an integer number of RUs. In Session 1 (the primary transmission session) the PT transmits the primary message $\Ip$ using a certain fraction of the $B$ RUs, while the PR, the ST and the SR are receiving, and feeding back Ack/Nacks. The PT continues transmitting until all packets from $\Ip$ have been received successfully by either the PR or the ST jointly, characterized by $\widehat{\mathcal{I}}_p \cup \Q_p = \Ip$. Given that the ST cooperates as a relay to assist in delivering $\Ip$, the remaining RUs are granted to the transmission of the ST. 
In Session 2 (the secondary transmission session) and Session 3 (the retransmission session) the ST takes on the role as a relay for both systems and retransmits all lost primary and secondary packets from the previous sessions, using one of three retransmission strategies described in Subsection \ref{s2-1}.

We consider two philosophically different constraints on the \emph{instantaneous frame size} $B$. In the first case we require that all primary and secondary packets be successfully received. Therefore, $B \leq \infty$ is determined as the total number of packet transmissions required for successful reception of all packets. It follows that the instantaneous frame size is a random variable $\boldsymbol{B}$ with the probability mass function $P_{\boldsymbol{B}}(B)$, where the frame size is adapted to the prevailing transmission conditions. Here $P_{\boldsymbol{B}}(B)=0$ for $B < N_p + N_s$. In this case there is no packet loss as the size of each frame is adapted to allow for successful reception. However, there is a non-zero probability that the reception delay may be excessively large as $B \leq \infty$. This case of no-packet-loss is therefore mainly of theoretical interest, and is referred to as the \emph{adaptive frame-size} (afs) case. To avoid large reception delays, we restrict the instantaneous frame size in the second case to be no larger than $\widehat{B}$; in other words $B \leq \widehat{B}$. However, there is now a non-zero probability that we are not able to successfully receive all primary and secondary data packets within a frame. Such an unsuccessful frame is defined as being lost due to a frame outage, and therefore the system is associated with a certain frame outage probability $P_{\text{out}}(B > \widehat{B})$. We refer to this case as the \emph{truncated frame-size} (tfs) case.

\subsection{Retransmission Strategies}\label{s2-1}
As mentioned earlier, we consider the conventional ARQ scheme as a baseline strategy. To improve the overall throughput, by providing additional cooperative throughput gain, we further consider the two network-coding schemes considered in our prior work \cite{LXR_VTC14}. Transmission Session 1, as described above, is the same for all three schemes. Transmission Session 2 is the same for the two network-coding schemes, but different for the conventional ARQ scheme. Session 3 (the retransmission session) is different for all three cases.

\begin{figure*}[t]
\centering
\subfigure[$\txm{4}\oplus\txm{f}$, $\txm{10}\oplus\txm{h}$, 6; $N_p=N_s=10$.]
{\includegraphics[width=70mm]{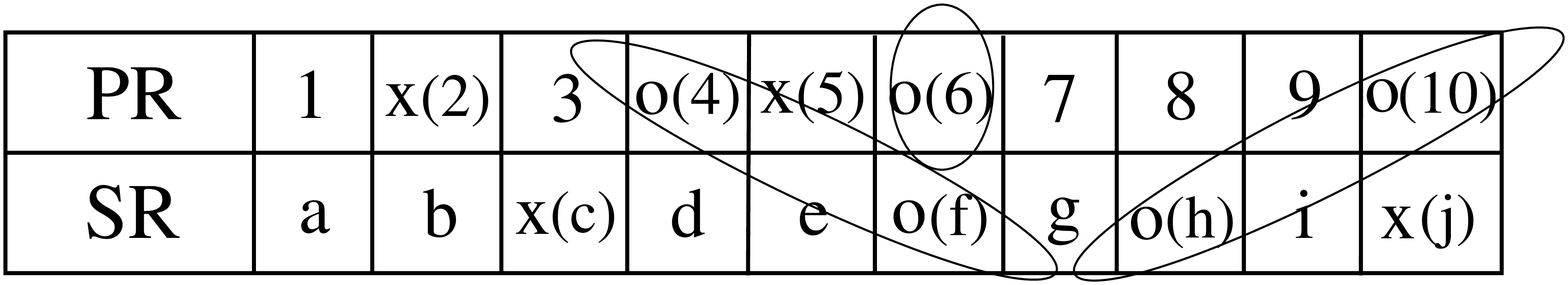}
\label{fig3:snc}} \hfil
\centering
\subfigure[$\txm{4}\oplus\txm{f}$, $\txm{6}\oplus\txm{f}$, $\txm{10}\oplus\txm{h}$; $N_p=N_s=10$.]
{\includegraphics[width=70mm]{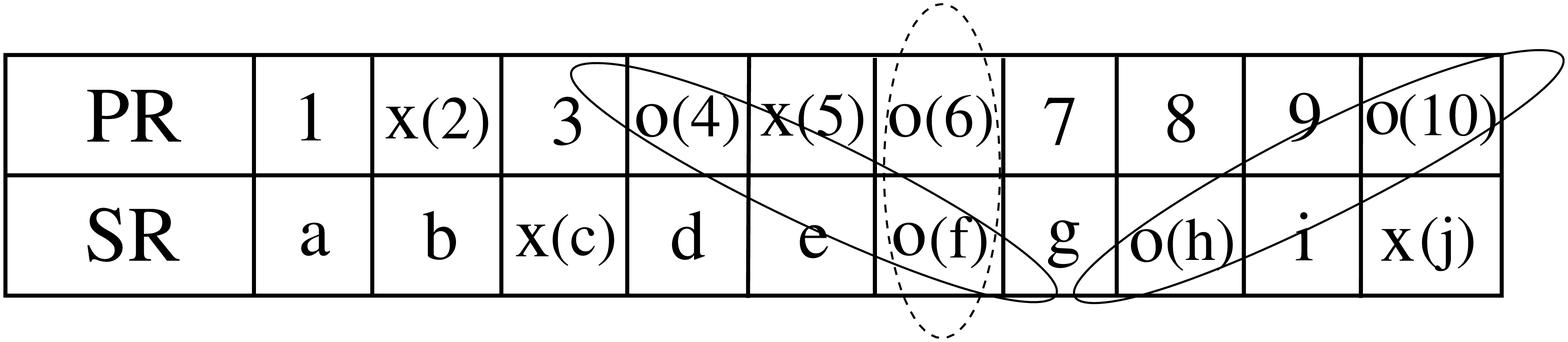}
\label{fig3:anc}} \hfil
\caption{Combined packets for the SNC scheme and ANC scheme,  respectively, in Fig. 3(a) and 3(b).}
\end{figure*}

\vspace{2pt}
\subsubsection{Conventional ARQ}
In Session 2 the ST relays all the packets lost at the PR until all primary packets have been successfully received. Here the ST gives strict priority to the primary packets and relays them before its own initial transmission. Subsequently, in Session 3 the ST transmits its own packets to the SR until all secondary packets have been successfully received.

\vspace{2pt}
\subsubsection{Static Network Coding (SNC)}
To enable network coding in Session 3, the ST will transmit the secondary message $\Is$ with no retransmissions in Session 2, while both the PR and the SR are receiving and feeding back Ack/Nacks. It follows that $B^\txs{s}_2 = B^\txs{a}_2 = N_s$. During Session 3, the retransmission session, the ST generates a sequence of as many new packets as possible by \textsc{xor}-ing a primary packet lost at the PR but received at the SR with a secondary packet lost at the SR but received at the PR. More formally, a coded packet is formed by combining a packet from $\overline{\Q}_p \cap \Pp$ with a packet from $\overline{\Q}_s \cap \Ps$. 
This sequence of coded packets is then transmitted to the two receivers. The PR (SR) is able to recover its lost packets since the secondary (primary) packets involved in the coding process are known at the PR (SR).

The combined packets may get lost during retransmission, thus triggering yet another retransmission. The ST will keep retransmitting a combined packet until it is successfully received at both receivers. Once all coded packets have been successfully delivered, the packets left in $\overline{\Q}_p$ and in $\overline{\Q}_s$ are transmitted individually, as for the conventional ARQ scheme, to the PR and the SR. A pattern of lost packets for the PR and the SR is shown in Fig. \ref{fig3:snc}. All the lost packets are denoted by circles (o) and crosses (x), in which ``o'' indicates the packet lost at the corresponding receiver but received by the other one, whereas ``x'' indicates the packet lost at both receivers. Obviously, only the ``o'' packets can be network coded. The combined packets are $\txm{4}\oplus\txm{f}$ and $\txm{10}\oplus\txm{h}$. The PR recovers packet 4 by $\txm{f}\oplus(\txm{4}\oplus\txm{f})$ and packet 10 by $\txm{h}\oplus(\txm{10}\oplus\txm{h})$; the SR recovers packet f by $\txm{4}\oplus(\txm{4}\oplus\txm{f})$ and packet h by $\txm{10}\oplus(\txm{10}\oplus\txm{h})$.

\vspace{2pt}
\subsubsection{Adaptive Network Coding (ANC)}
From the description above, it is clear that Session 3 in the SNC scheme is sub-optimal since the ST is required to retransmit the same coded packet even if one of the receivers has successfully recovered one of the involved packets. Instead, the ST can dynamically form a new packet by \textsc{xor}-ing the un-recovered packet with one of the packets left in the encodable packet set. With reference to Fig. \ref{fig3:anc}, suppose that packet $\txm{4}\oplus\txm{f}$ is received at the PR but lost at the SR. In the next transmission attempt the ST transmits $\txm{6}\oplus\txm{f}$ instead of $\txm{4}\oplus\txm{f}$. The number of transmission attempts using ANC is therefore generally reduced as compared to SNC.

\subsection{Performance Metrics}\label{s2-2}
In wireless networks, there are many important performance metrics. Here our main focus is on throughput-delay/outage-probability tradeoffs and their relationship with our two constraints on the instantaneous frame size.
For a cognitive radio network, we typically define the throughput for the primary system and secondary system separately by $\eta_p$ and $\eta_s$, as the average number of packets that are successfully delivered per resource unit in each system. With the assumption of an adaptive frame size we have the throughputs as:
\begin{align}
&\eta_p^{\text{afs}} = \frac{N_p}{\overline{B}_{\text{afs}}}, \hspace{8mm} \eta_s^{\text{afs}} = \frac{N_s}{\overline{B}_{\text{afs}}}, \label{eq:eta1}
\end{align}
where
\begin{align}
&\overline{B}_{\text{afs}} = E_{\boldsymbol{B}}[B\;|\;B\leq\infty] = \sum_{B=1}^{\infty} B \cdot P_{\boldsymbol{B}}(B).
\end{align}
With the assumption of a truncated frame size we have the throughputs as:
\begin{align}
&\eta_p^{\text{tfs}} = \frac{N_p}{\overline{B}_{\text{tfs}}}, \hspace{8mm} \eta_s^{\text{tfs}} = \frac{N_s} {\overline{B}_{\text{tfs}}}, \label{eq:eta2}
\end{align}
where
\begin{align}
&\overline{B}_{\text{tfs}}  = E_{\boldsymbol{B}}[B\;|\;B\leq\widehat{B}] =\frac{1}{1 - P_{\text{out}}(\widehat{B})}\sum_{B=1}^{\widehat{B}} B \cdot  P_{\boldsymbol{B}}(B),
\end{align}
and the outage probability is determined as
\begin{align}
P_{\text{out}}(\widehat{B})&=\mathbb{P}\{B> \widehat{B}\} = \sum_{B=\widehat{B}+1}^{\infty}P_{\boldsymbol{B}}(B). \label{eq:outage}
\end{align}

We analyze the throughput and outage performance of the three transmission strategies in the following two sections. We first consider the analysis of the throughput performance for the adaptive frame-size case. As clear from the performance metrics, the task is therefore to determine the average frame size, subject to each of the three transmission strategies. We subsequently leverage this analysis to determine the throughput and outage performances for the truncated frame-size case.


\section{Performance Analysis for Adaptive Frame Size}\label{sec:performance_a}

Here, we analyze the throughput performance of the three transmission strategies outlined above for the adaptive frame size case.
As previously defined, each transmission frame is divided into three sessions, the individual primary and secondary transmission sessions, and the retransmission session. For each session, we denote by $B_1$, $B_2$ and $B_3$ the instantaneous number of transmissions realised in each respective session, where $B = B_1+B_2+B_3$ is the total number of transmissions.
We will first consider the conventional ARQ scheme for the primary-secondary cooperation.

\subsection{Conventional ARQ Scheme}
In the first session the PT keeps transmitting until the $N_p$ primary packets have been received by either the PR or the ST. This is a simple case of conventional ARQ over a PEL with a packet loss probability of $p_1q$. The number of transmissions required by the PT follows a negative binomial distribution, namely $B^\txs{c}_1 \sim \mathcal{NB}(B,N_p,p_1q)$, as argued in \cite{Larsson08wcnc}. Similarly, in the second session the number of retransmissions $B^\txs{c}_2$ follows a negative binomial distribution, depending on the number of packets lost by the PR. We denote the number of lost primary packets by $L_p$, corresponding to $N_p - L_p$ packets received by the PR. In the third session of secondary transmission, the number of retransmissions $B^\txs{c}_3$ also follows a negative binomial distribution as $B^\txs{c}_3 \sim \mathcal{NB}(B,N_s,p_2)$. The probability mass functions of the number of transmissions for each session are defined as:
\begin{subequations}
\begin{align}
&\mathbb{P}\figbrc{B^\txs{c}_1=B_1} = \binom{B_1-1}{N_p-1}(p_1q)^{B_1-N_p}(1-p_1q)^{N_p} \label{eq:Pn1c}\\
&\mathbb{P}\figbrc{B^\txs{c}_2=B_2\mid L_p=k_p} = \binom{B_2-1}{k_p-1}p_{21}^{B_2-k_p}(1-p_{21})^{k_p} \label{eq:Pn2c}\\
&\mathbb{P}\figbrc{B^\txs{c}_3=B_3} = \binom{B_3-1}{N_s-1}p_2^{B_3-N_s}(1-p_2)^{N_s}, \label{eq:Pn3c}
\end{align}
\end{subequations}
in which the number of lost packets $L_p $ at the PR is a binomial distributed random variable, with the probability mass function
\begin{align}
&\mathbb{P}\figbrc{L_p=k_p} = \binom{N_p}{k_p}\brc{\frac{p_1(1-q)}{1-p_1q}}^{k_p}\brc{1-\frac{p_1(1-q)}{1-p_1q}}^{N_p-k_p}. \label{eq:Pkc}
\end{align}
The unconditional probability mass function of $B^\txs{c}_2$ can be determined jointly by \eqref{eq:Pn2c} and \eqref{eq:Pkc} as
\begin{equation*}
\mathbb{P}\figbrc{B^\txs{c}_2=B_2} = \sum^{N_p}_{k_p=0} \mathbb{P}\figbrc{L_p=k_p}\mathbb{P}\figbrc{B^\txs{c}_2=B_2\mid L_p=k_p}.
\end{equation*}
As the three sessions operate independently of each other in terms of the number of packets transmitted, the expected frame size is determined as:
\begin{align}
\overline{B}^\txs{c}_{\text{afs}} &= E_{\boldsymbol{B}}[B^\txs{c}_{1} + B^\txs{c}_{2} + B^\txs{c}_{3}\;|\;B\leq\infty] \nonumber \\
&= E_{\boldsymbol{B}}[B^\txs{c}_{1}\;|\;B\leq\infty] + E_{\boldsymbol{B}}[B^\txs{c}_{2}\;|\;B\leq\infty] + E_{\boldsymbol{B}}[B^\txs{c}_{3}\;|\;B\leq\infty] \nonumber \\
&= \frac{N_p}{1-p_1q} + \frac{N_pp_1(1-q)}{(1-p_1q)(1-p_{21})} + \frac{N_s}{1-p_2}. \label{eq:afs_c}
\end{align}
Even though we can determine the expected frame size, the sum of negative binomial random variables is not necessarily negative binomial distributed. So for the analysis of the two remaining schemes, as well as for the truncated frame size, we consider the following Lemma to obtained a tractable analytical framework.
\begin{lemma}\label{Lemma1}
As $B$ and $N$ increases and with $\delta < p < 1-\gamma$ for appropriately small $\delta$ and $\gamma$, the negative binomial distribution $\mathcal{NB}(B,N,p)$ approaches a normal distribution $\mathcal{N}(\mu,\sigma)$, where $\mu = N/(1-p)$ and $\sigma=\sqrt{Np/(1-p)^2}$. (For proof: See \cite{B66}.)\footnote{Since the frame size is limited to $(N_p +N_s \leq B \leq \infty)$, a lower-truncated normal distribution \cite{JB14,CUD94} may be a better approximation; however, when $(N_p+N_s)$ is sufficiently large, the probability of frame sizes smaller than $(N_p+N_s)$ is negligible. We therefore consider a standard normal distribution to allow for notational clarity. In Section \ref{sec:performance_p} we consider the use of an upper-truncated normal distribution to analyze the performance for an upper-bounded frame size.}
\end{lemma}
Here we assume that the packet loss probability of each link is neither too large nor too small. Thus the number of transmissions in each session can be approximated as a normal distributed random variable when $N_p$ and $N_s$ are sufficiently large. Furthermore, as the transmissions in different sessions are independent from each other, the total number of transmissions is also normal distributed with additive mean and variance \cite{PM_PB95}. For the conventional ARQ scheme, the number of transmissions in each session can be approximated by a normal distribution with mean and standard deviation as follows:
\begin{subequations}
\begin{align}
&B^\txs{c}_1 \sim \mathcal{N}\brc{\mu^\txs{c}_1 = \frac{N_p}{1-p_1q},\; \sigma^\txs{c}_1 = \sqrt{\frac{\mu^\txs{c}_1p_1q}{1-p_1q}}} \label{eq:arq_1}\\
&B^\txs{c}_2 \sim \mathcal{N}\brc{\mu^\txs{c}_2 = \frac{N_pp_1(1-q)}{(1-p_1q)(1-p_{21})},\; \sigma^\txs{c}_2 = \sqrt{\frac{\mu^\txs{c}_2p_{21}}{1-p_{21}}}} \label{eq:arq_2} \\
&B^\txs{c}_3 \sim \mathcal{N}\brc{\mu^\txs{c}_3 = \frac{N_s}{1-p_2},\; \sigma^\txs{c}_3 = \sqrt{\frac{\mu^\txs{c}_3p_2}{1-p_2}}}, \label{eq:arq_3}
\end{align}
\end{subequations}
which makes the total number of transmissions $B^\txs{c} \sim \mathcal{N}\figbrc{\mu^\txs{c},\sigma^\txs{c}}$ with $\mu^\txs{c}=\mu^\txs{c}_1+\mu^\txs{c}_2+\mu^\txs{c}_3$ and $(\sigma^\txs{c})^2=(\sigma^\txs{c}_1)^2+(\sigma^\txs{c}_2)^2+(\sigma^\txs{c}_3)^2$. We note that the expected frame size is the same as in (\ref{eq:afs_c}).

\subsection{Stationary Network Coding (SNC) Scheme}

For the SNC scheme, Session 1 is the same as for conventional ARQ, and thus $B^\txs{s}_1 \sim \mathcal{NB}(B,N_p,p_1q)$ (approximated by (\ref{eq:arq_1})). Furthermore, Session 2 is just to forward all secondary packets without any retransmissions, and thus $B^\txs{s}_2 = N_s$. Therefore, to determine the total average number of transmissions, we only need to determine the average number of transmissions in Session 3. Let $L_p$ and $L_s$ be the number of lost packets at each receiver after the first two sessions, i.e., $L_p=|\overline{\Q}_p|=N_p-|\Q_p|$ and $L_s=|\overline{\Q}_s|=N_s-|\Q_s|$. The probability that the PR has lost $k_p$ packets and the SR has lost $k_s$ packets is determined as
\begin{align}
&\mathbb{P}\figbrc{L_p=k_p,L_s=k_s} = \binom{N_p}{k_p}\brc{\frac{p_1(1-q)}{1-p_1q}}^{k_p}\brc{1-\frac{p_1(1-q)}{1-p_1q}}^{N_p-k_p} \cdot \binom{N_s}{k_s}p^{k_s}_2(1-p_2)^{N_s-k_s}. \label{eq:Pks}
\end{align}

The lost packets can be divided into three subsets for retransmission; the network-coded packets defined by the set $\mathcal{C}$, where the number of possible network-coded packets is the minimum of $|\overline{\Q}_p \cap \Pp|$ and $|\overline{\Q}_s \cap \Ps|$. We further denote $k_{\min}$ as $|\mathcal{C}|$, determined as $\min \figbrc{|\overline{\Q}_p \cap \Pp|, |\overline{\Q}_s \cap \Ps|}$. The remaining primary packets in $\overline{\Q}_p\backslash\mathcal{C}$ and secondary packets in $\overline{\Q}_s\backslash\mathcal{C}$  are to be transmitted separately to the PR and the SR, respectively, using conventional ARQ.

Given that $k_p$ primary and $k_s$ secondary packets are lost, the conditional probability of the number of retransmissions of the SNC scheme is determined as
\begin{align}
&\mathbb{P}\figbrc{B^\txs{s}_3 = B_3\mid L_p=k_p,L_s=k_s}
= \mathbb{P}\figbrc{B^\txs{s}_3(\mathcal{C})+B^\txs{s}_3(\overline{\Q}_p\backslash\mathcal{C})+B^\txs{s}_3(\overline{\Q}_s\backslash\mathcal{C}) \mid L_p=k_p,L_s=k_s}, \label{eq:CondPn3s}
\end{align}
where $B^\txs{s}_3(\mathcal{C})$ is the number of transmissions of network-coded packets from the set $\mathcal{C}$. $B^\txs{s}_3(\overline{\Q}_p\backslash\mathcal{C})$ and $B^\txs{s}_3(\overline{\Q}_s\backslash\mathcal{C})$ denote the number of individually retransmitted primary and secondary packets to the corresponding receivers. The unconditional probability is determined as
\begin{align} \label{eq:Pb3s}
&\mathbb{P}\figbrc{B^\txs{s}_3 = B_3}
= \\ \nonumber
&\sum_{k_p=0}^{N_p} \sum_{k_s=0}^{N_s}\mathbb{P}\figbrc{B^\txs{s}_3(\mathcal{C})+B^\txs{s}_3(\overline{\Q}_p\backslash\mathcal{C})+B^\txs{s}_3(\overline{\Q}_s\backslash\mathcal{C}) \mid L_p=k_p,L_s=k_s}  \mathbb{P}\figbrc{L_p=k_p,L_s=k_s},
\end{align}
from which the expected number of transmission attempts for the retransmission session can be determined. However, as the unconditional probability of $B^\txs{s}_3$ is computationally challenging, we apply instead the law of total expectation \cite{NW06} to derive the conditional expected value of $B^\txs{s}_3$ as
\begin{align}
&E_{\boldsymbol{B}}[B^\txs{s}_{3}\;|\;B\leq\infty] = E_{\{L_p,L_s\}}\sqbrc{E_{\boldsymbol{B}}[B^\textsc{s}_3\mid L_p=k_p,L_s=k_s]} \label{eq:En3s} \\
= &\sum^{N_p}_{k_p=0}\sum^{N_s}_{k_s=0}{\mathbb{P}\figbrc{L_p=k_p,L_s=k_s}E_{\boldsymbol{B}}[B^\textsc{s}_3\mid L_p=k_p,L_s=k_s]}. \nonumber
\end{align}
Here $\mathbb{P}\figbrc{L_p=k_p,L_s=k_s}$ is given in \eqref{eq:Pks}. In our case, the retransmission process of the network-coded packets from the ST to the PR and the SR is considered as a two-receiver broadcast process, where each packet should be received successfully by both receivers. Thus the transmission efficiency of the network-coded packets is $\mu^\textsc{bc}(2) = \frac{1}{1-p_{21}} + \frac{1}{1-p_2} - \frac{1}{1-p_2p_{21}}$ (For proof: See Appendix \ref{App1}). When there are $k_{\min}$ network-coded packets to be transmitted, the expected number of transmissions to ensure that both receivers successfully receive these packets is simply $k_{\min}\mu^{\txm{BC}}(2)$. The expectation in the double summation is therefore given by
\begin{align}
&E_{\boldsymbol{B}}[B^\textsc{s}_3\mid L_p=k_p,L_s=k_s] \label{eq:CondEn3s}\\
= &\sum^{k_p}_{i=0}\sum^{k_s}_{j=0}\binom{k_p}{i}(1-p_{12})^ip^{k_p-i}_{12} \cdot \binom{k_s}{j}(1-p_{21})^jp^{k_s-j}_{21} \nonumber\\
&\cdot \brc{k_{\min}\mu^{\txm{BC}}(2) + \frac{k_p-k_{\min}}{1-p_{21}} + \frac{k_s-k_{\min}}{1-p_2}}, \nonumber
\end{align}
where $k_{\min}=\min\{i,j\}$, which indicates the maximum number of possible coded packets the ST can transmit by matching pairs of lost packets in $L_p$ and $L_s$; and $\mu^\textsc{bc}(2)$ is the average number of transmission attempts per packet for a two-receiver broadcast channel. 

For the two subsets, $\overline{\Q}_p\backslash\mathcal{C}$ and $\overline{\Q}_s\backslash\mathcal{C}$, respectively, the transmission processes can be characterised by appropriate negative binomial distributions; however, for the network-coded packet subset $\mathcal{C}$, the transmission process is characterised as a random variable $B^\txs{s}_3(\mathcal{C})$. For each of the packets in $\mathcal{C}$, the transmission is characterised by the maximum of two independent negative binomial random variables. For $k_{\min}=|\mathcal{C}|$ packets, the transmission process needs to be repeated for $k_{\min}$ times. Moreover, the sum over $k_p$ and $k_s$ of these random variables are not simply negative binomial distributed. We therefore again consider the use of Lemma \ref{Lemma1}.

The number of transmissions for delivering $k_{\min}$ packets is a random variable $B^\txs{s}_3(\mathcal{C})=\sum^{k_{\min}}_{n=1}\max\{X_{1,n},X_{2,n}\}$, where $X_{1,n}$ and $X_{2,n}$ represent the numbers of transmission attempts for delivering the $n$th packet to the PR and the SR, respectively. Both $X_{1,n}$ and $X_{2,n}$ are negative binomial distributed with the probabilities
\begin{subequations}
\begin{empheq}[left=\empheqlbrace]{align}
\mathbb{P}\figbrc{X_{1,n}=m_1} &= p^{m_1-1}_{21}(1-p_{21}), \;\;\; \text{for}\;\; m_1=1,2,... \label{eq:X1s}\\
\mathbb{P}\figbrc{X_{2,n}=m_2} &= p^{m_2-1}_2(1-p_2), \;\;\; \text{for}\;\; m_2=1,2,... \label{eq:X2s}
\end{empheq}
\end{subequations}
and by Lemma \ref{Lemma1}, they can be approximated by normal distributions with mean and standard deviation as
\begin{subequations}
\begin{empheq}[left=\empheqlbrace]{align}
&X_{1,n} \sim \mathcal{N}\brc{\mu_{X_{1,n}}=\frac{1}{1-p_{21}},\; \sigma_{X_{1,n}}=\sqrt{\frac{p_{21}}{(1-p_{21})^2}}} \nonumber\\
&X_{2,n} \sim \mathcal{N}\brc{\mu_{X_{2,n}}=\frac{1}{1-p_2},\; \sigma_{X_{2,n}}=\sqrt{\frac{p_2}{(1-p_2)^2}}}. \nonumber
\end{empheq}
\end{subequations}

\begin{figure*}
\centering
\hspace{-0.36cm}
\subfigure[$p_{21}=0.1$, $p_2=0.1$]{
\includegraphics[width=0.39\textwidth,height=5.0cm]{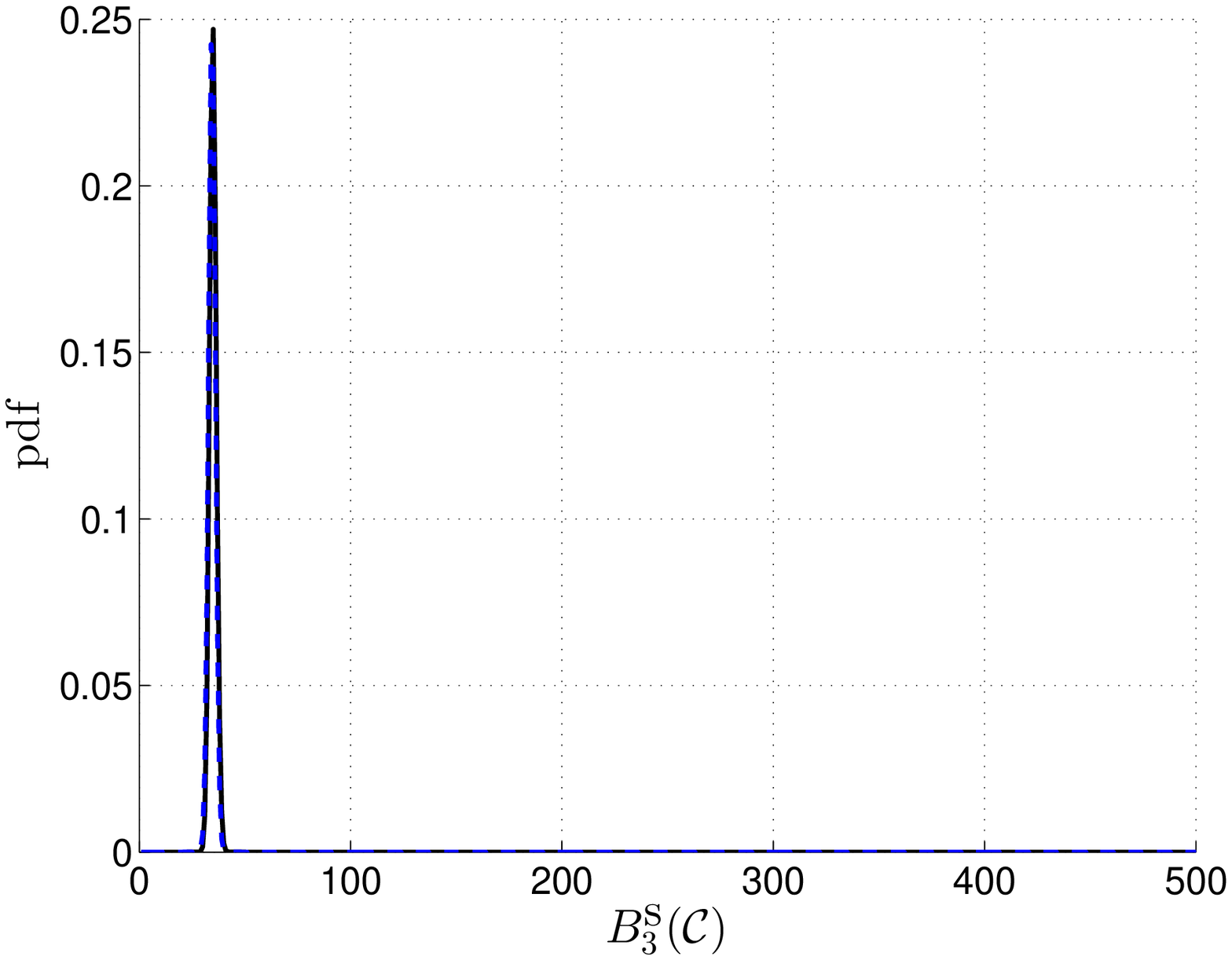}
} \hfil 
\subfigure[$p_{21}=0.1$, $p_2=0.5$]{
\includegraphics[width=0.37\textwidth,height=5.0cm]{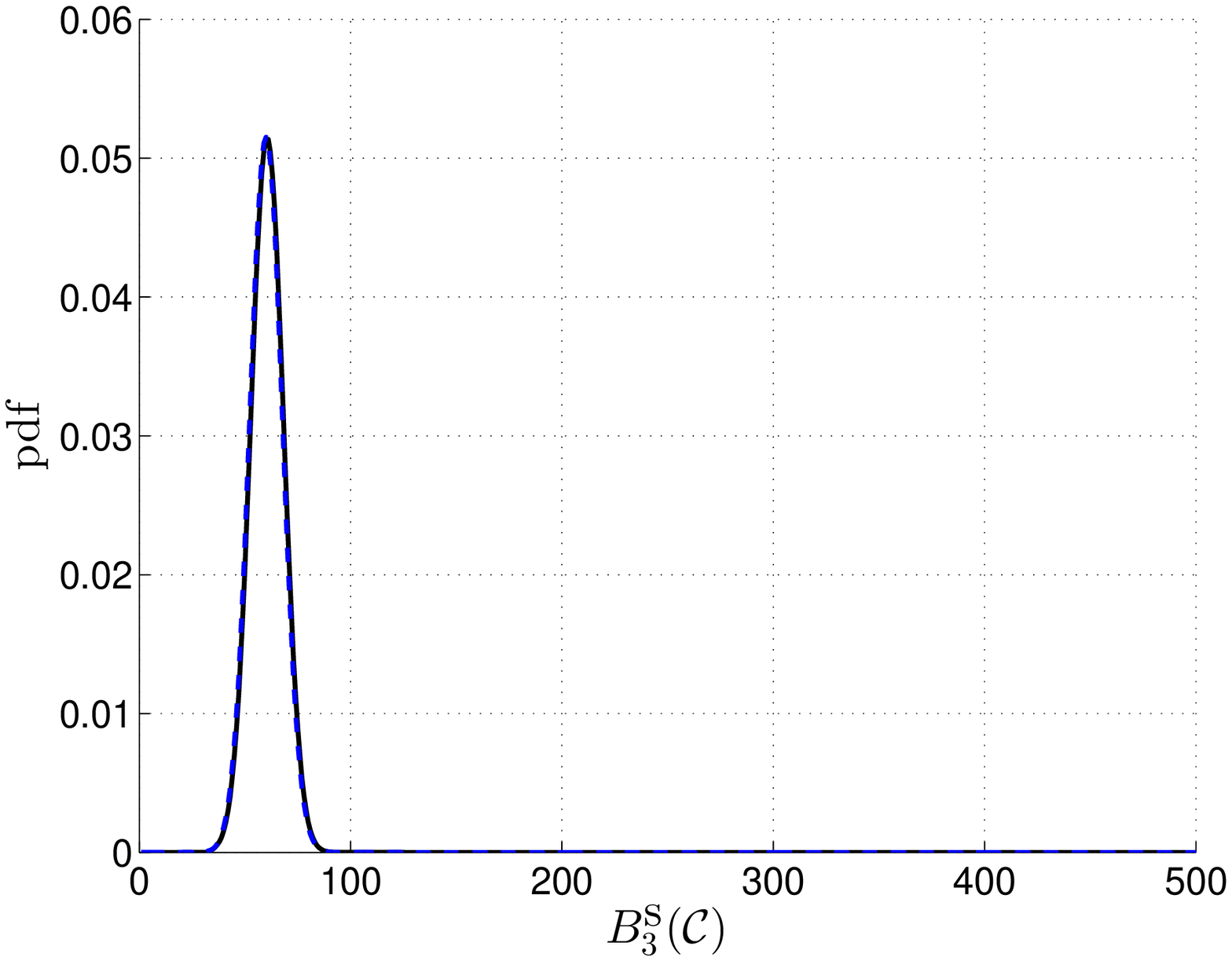}
}
\subfigure[$p_{21}=0.1$, $p_2=0.9$]{
\includegraphics[width=0.37\textwidth,height=5.0cm]{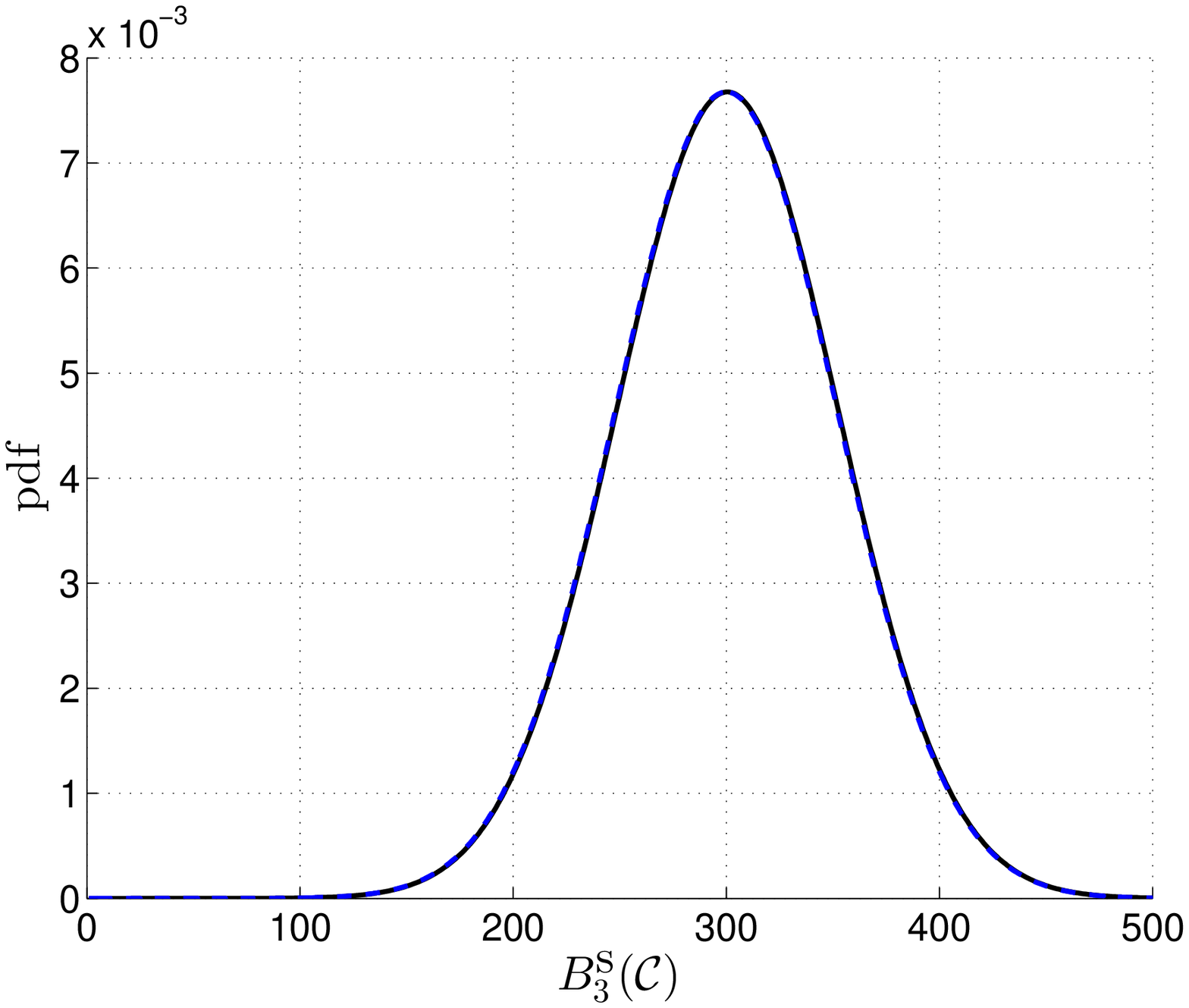}
} \hfil
\subfigure[$p_{21}=0.9$, $p_2=0.9$]{
\includegraphics[width=0.37\textwidth,height=5.0cm]{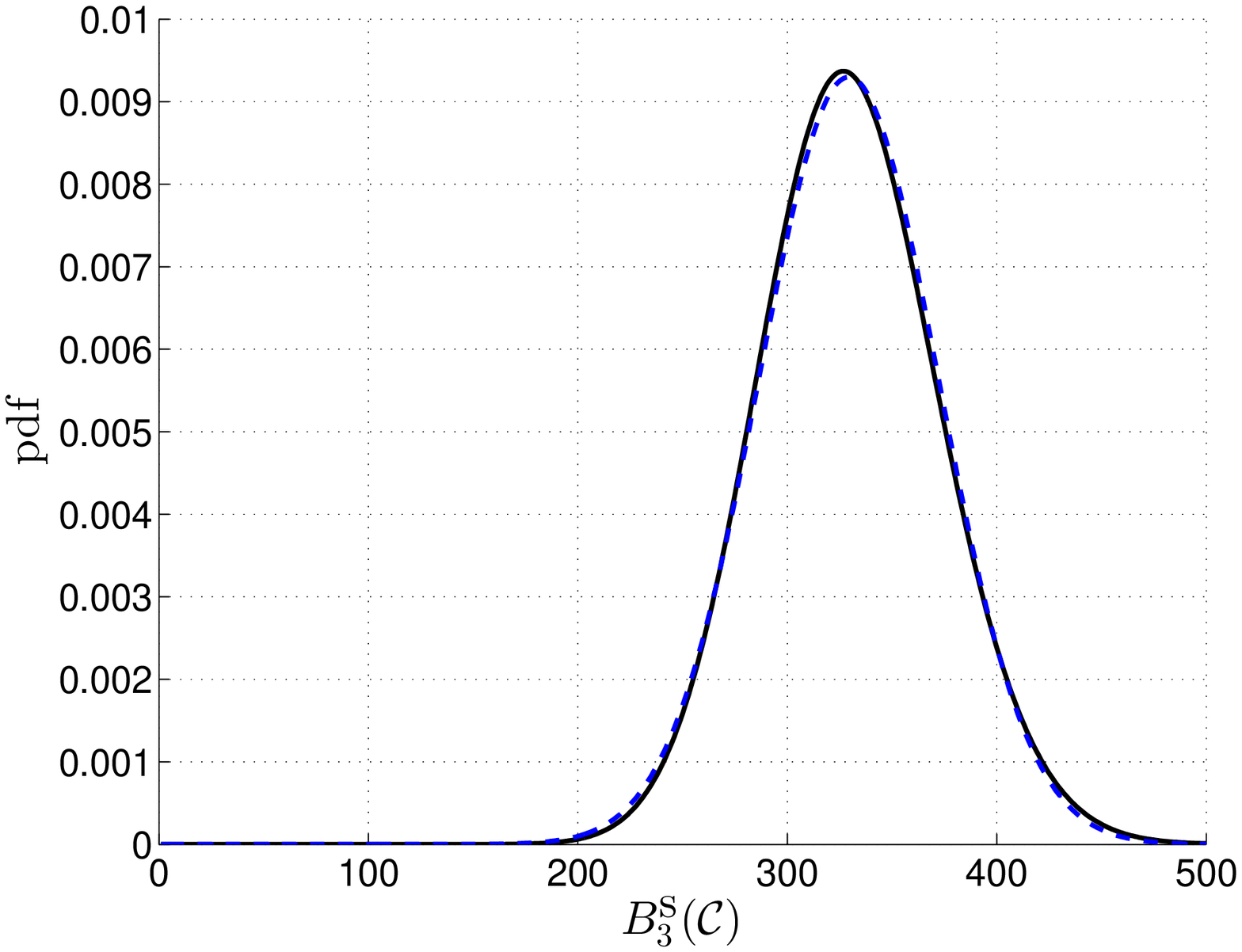}
}
\caption{Normal approximation to the distribution of $\max\{X_1,X_2\}$.}
\label{fig:max_normal}
\end{figure*}

For the distribution of the maximum/minimum of two independent normally distributed random variables, the moments of order statistics was determined in the 1950's \cite{C61}. Numerical results show that when the difference between the standard deviations is small, the distribution of the maximum is well approximated by a normal distribution \cite{NK08}. In our case, the standard deviation of $X_{1,n}$ and $X_{2,n}$ is related to the corresponding link quality, with the packet erasure probability $p_{21}$ and $p_2$. Thus we show the normal approximation to $\max\{X_{1,n},X_{2,n}\}$ as a function of the link qualities when $k_{\min}=30$ in Fig. \ref{fig:max_normal}. Without loss of generality, considering the possible range of values of $p_{21}$ and $p_2$, we compare the pdf of $\max\{X_{1,n},X_{2,n}\}$, denoted by the solid line, with its normal approximation, denoted by the dashed line, in four cases: (a) $p_{21}=0.1$ and $p_2=0.1$; (b) $p_{21}=0.1$ and $p_2=0.5$; (c) $p_{21}=0.1$ and $p_2=0.9$; and (d) $p_{21}=0.9$ and $p_2=0.9$. For all cases, the normal approximation matches the practical pdf almost perfectly. Since the packet erasure probabilities are constraint to $0\leq p\leq 1$, the deviation between the approximation and the practical pdf can be neglected. We approximate $\max\{X_{1,n},X_{2,n}\}$ by a normal distribution with mean and standard deviation as
\begin{align} \label{eq:max_normal}
\mu_{\max}(n) = &\mu_{X_{1,n}}\Phi\brc{\frac{\mu_{X_{1,n}}-\mu_{X_{2,n}}}{\theta}} + \mu_{X_{2,n}}\Phi\brc{\frac{\mu_{X_{2,n}}-\mu_{X_{1,n}}}{\theta}} + \theta\phi\brc{\frac{\mu_{X_{1,n}}-\mu_{X_{2,n}}}{\theta}} \nonumber\\
\sigma^2_{\max}(n) = &\brc{\sigma^2_{X_{1,n}}+\mu^2_{X_{1,n}}}\Phi\brc{\frac{\mu_{X_{1,n}}-\mu_{X_{2,n}}}{\theta}} + \brc{\sigma^2_{X_{2,n}}+\mu^2_{X_{2,n}}}\Phi\brc{\frac{\mu_{X_{2,n}}-\mu_{X_{1,n}}}{\theta}} \nonumber\\
&+ \brc{\mu_{X_{1,n}}+\mu_{X_{2,n}}}\theta\phi\brc{\frac{\mu_{X_{1,n}}-\mu_{X_{2,n}}}{\theta}} - \mu^2_{\max}(n),
\end{align}
where $\Phi$ and $\phi$ denote the the cumulative probability function (cdf) and the probability density function (pdf) of the standard normal distribution respectively and $\theta = \sqrt{\sigma^2_{X_{1,n}}+\sigma^2_{X_{2,n}}}$.
Since the transmission for each packet is independent, $B^\txs{s}_3(\mathcal{C})$, as the sum of $k_{\min}$ normal distributed random variables, is also normally distributed with $\mu\brc{B^\txs{s}_3(\mathcal{C})} = \sum^{k_{\min}}_{n=1}\mu_{\max}(n)$ and $\sigma^2\brc{B^\txs{s}_3(\mathcal{C})} = \sum^{k_{\min}}_{n=1}\sigma^2_{\max}(n)$.

Based on the analysis above, the number of retransmissions for each subset can be found by approximating the conditional probability shown in \eqref{eq:CondPn3s} by a normal distribution with the moment parameters in \eqref{eq:dist_n3s} below.
\begin{subequations} \label{eq:dist_n3s}
\begin{eqnarray}
B^\txs{s}_3(\mathcal{C}) &\sim& \mathcal{N}\brc{\mu\brc{B^\txs{s}_3(\mathcal{C})},\; \sigma\brc{B^\txs{s}_3(\mathcal{C})}} \\
B^\txs{s}_3(\overline{\Q}_p\backslash\mathcal{C}) &\sim& \mathcal{N}\brc{\frac{\frac{N_pp_1(1-q)}{1-p_1q}-k_{\min}}{1-p_{21}},\; \sqrt{\brc{\frac{N_pp_1(1-q)}{1-p_1q}-k_{\min}}\cdot \frac{p_{21}}{(1-p_{21})^2}}} \\
B^\txs{s}_3(\overline{\Q}_s\backslash\mathcal{C}) &\sim& \mathcal{N}\brc{\frac{N_sp_2-k_{\min}}{1-p_2},\; \sqrt{\brc{N_sp_2-k_{\min}}\cdot \frac{p_2}{(1-p_2)^2}}}
\end{eqnarray}
\end{subequations}

Moreover, $k_{\min}$ in \eqref{eq:dist_n3s} above is a random variable denoting the maximum number of possible coded packets and $k_{\min}=|\mathcal{C}|=\min\{|\overline{\Q}_p \cap \Pp|, |\overline{\Q}_s \cap \Ps|\}$. To derive the moment parameters, we decide to apply the mean value of $k_{\min}$. As the number of the encodable packets in $\overline{\Q}_p$ and $\overline{\Q}_s$ is binomial distributed conditioned on $k_p$ and $k_s$,
\begin{subequations}
\begin{align}
&\mathbb{P}\figbrc{|\overline{\Q}_p \cap \Pp|=i} = \binom{k_p}{i}(1-p_{12})^ip^{k_p-i}_{12}, \;\;\; \text{for}\;\; i=0,1,... \nonumber\\
&\mathbb{P}\figbrc{|\overline{\Q}_s \cap \Ps|=j} = \binom{k_s}{j}(1-p_{21})^jp^{k_s-j}_{21}, \;\;\; \text{for}\;\; j=0,1,..., \nonumber
\end{align}
\end{subequations}
which can be approximated by normal distributions
\begin{subequations}
\begin{align}
&|\overline{\Q}_p \cap \Pp| \sim \mathcal{N}\brc{\mu_i=k_p(1-p_{12}),\; \sigma_i=\sqrt{k_pp_{12}(1-p_{12})}} \nonumber\\
&|\overline{\Q}_s \cap \Ps| \sim \mathcal{N}\brc{\mu_j=k_s(1-p_{21}),\; \sigma_j=\sqrt{k_sp_{21}(1-p_{21})}}. \nonumber
\end{align}
\end{subequations}
As we mentioned above, for the distribution of the minimum of two independent normally distributed random variables, the moments of order statistics can be determined. Thus, the mean value of $k_{\min}$ can be derived as
\begin{align}
&\mu(k_{\min}) = \mu_i\Phi\brc{\frac{\mu_j-\mu_i}{\theta_{k_{\min}}}} + \mu_j\Phi\brc{\frac{\mu_i-\mu_j}{\theta_{k_{\min}}}} - \theta_{k_{\min}}\phi\brc{\frac{\mu_i-\mu_j}{\theta_{k_{\min}}}}, \nonumber
\end{align}
where $\mu_i$ and $\mu_j$ are the mean value of the number of the encodable packets in each subset and $\theta_{k_{\min}} = \sqrt{\sigma^2_i+\sigma^2_j}$. We can substitute $k_{\min}$ by the result $\mu(k_{\min})$ in the approximation of $B^\txs{s}_3(\mathcal{C})$ in \eqref{eq:dist_n3s}.

\vspace{2pt}
\subsection{Adaptive Network Coding (ANC) Scheme}
In the retransmission session, for the ANC scheme, the ST dynamically forms another coded packet based on which receiver has received the previous one. Apart from this, the other transmission processes are the same as the SNC scheme. In contrast to the SNC scheme, here we define $L_p$ and $L_s$ as the number of packets that could be encoded by \textsc{xor}-ing, i.e., $L_p=|\overline{\Q}_p \cap \Pp|$ and $L_s=|\overline{\Q}_s \cap \Ps|$. The probability of $k_p$ encodable packets at the PR and $k_s$ at the SR is given by
\begin{eqnarray}
\mathbb{P}\figbrc{L_p=k_p,L_s=k_s} &=& \binom{N_p}{k_p}\brc{\frac{p_1(1-q)(1-p_{12})}{1-p_1q}}^{k_p}\brc{1-\frac{p_1(1-q)(1-p_{12})}{1-p_1q}}^{N_p-k_p} \nonumber\\
 &&\cdot \binom{N_s}{k_s}{\brc{p_2(1-p_{21})}}^{k_s}(1-p_2(1-p_{21}))^{N_s-k_s}.\label{eq:Pka}
\end{eqnarray}

In this case, all the required packets can be classified into three subsets: the encodable packets in $\overline{\Q}_p \cap \Pp$ and $\overline{\Q}_s \cap \Ps$, defined by the set $\mathcal{C}$, the individual primary packets in $\overline{\Q}_p \cap \overline{\mathcal{P}}_p$ and the individual secondary packets in $\overline{\Q}_s \cap \overline{\mathcal{P}}_s$ to be transmitted to the PR and the SR separately. Given $k_p$ and $k_s$ encodable packets at the PR and the SR, the conditional probability of the number of retransmissions of the ANC scheme is shown in \eqref{eq:CondPn3a}, where $B^\txs{a}_3(\mathcal{C})$ is the number of transmissions for all encodable packets. The unconditional probability is accordingly determined in \eqref{eq:Pn3a}.
\begin{align}
&\mathbb{P}\figbrc{B^\txs{a}_3 = B_3\mid L_p=k_p,L_s=k_s} = \mathbb{P}\figbrc{B^\txs{a}_3(\mathcal{C})+B^\txs{a}_3(\overline{\Q}_p \cap \overline{\mathcal{P}}_p)+B^\txs{a}_3(\overline{\Q}_s \cap \overline{\mathcal{P}}_s) \mid L_p=k_p,L_s=k_s} \label{eq:CondPn3a}
\end{align}
\begin{align} \label{eq:Pn3a}
&\mathbb{P}\figbrc{B^\txs{a}_3 = B_3}
= \\ \nonumber
&\sum_{k_p=0}^{N_p} \sum_{k_s=0}^{N_s}\mathbb{P}\figbrc{B^\txs{a}_3(\mathcal{C})+B^\txs{a}_3(\overline{\Q}_p \cap \overline{\mathcal{P}}_p)+B^\txs{a}_3(\overline{\Q}_s \cap \overline{\mathcal{P}}_s) \mid L_p=k_p,L_s=k_s}  \mathbb{P}\figbrc{L_p=k_p,L_s=k_s}.
\end{align}

The expected frame size for the ANC scheme $\overline{B}^\txs{a}_{\text{afs}}$ is similar to the SNC scheme, only with the expected number of transmission attempts for the retransmission session derived as
\begin{align}
&E_{\boldsymbol{B}}[B^\txs{a}_{3}\;|\;B\leq\infty] = E\sqbrc{E_{\boldsymbol{B}}[B^\textsc{a}_3\mid L_p=k_p,L_s=k_s]} \label{eq:En3a} \\
= &\sum^{N_p}_{k_p=0}\sum^{N_s}_{k_s=0}{\mathbb{P}\figbrc{L_p=k_p,L_s=k_s}E_{\boldsymbol{B}}[B^\textsc{a}_3\mid L_p=k_p,L_s=k_s]}, \nonumber
\end{align}
where $\mathbb{P}\figbrc{L_p=k_p,L_s=k_s}$ is given in \eqref{eq:Pka}. The expectation in the double summation is given by
\begin{align}
&E_{\boldsymbol{B}}[B^\txs{a}_3\mid L_p=k_p,L_s=k_s] \label{eq:CondEn3a}\\
= &\sum^{\infty}_{k=\max\{k_p,k_s\}}{k \mathbb{P}\figbrc{B^\txs{a}_3(\mathcal{C})=k}} + \frac{k_pp_{12}}{(1-p_{12})(1-p_{21})} + \frac{k_sp_{21}}{(1-p_{21})(1-p_2)}. \nonumber
\end{align}

The transmission processes for the two individually transmitted packet subsets, $\overline{\Q}_p \cap \overline{\mathcal{P}}_p$ and $\overline{\Q}_s \cap \overline{\mathcal{P}}_s$, are independently negative binomial distributed. Following Lemma \ref{Lemma1}, a normal distribution can be applied appropriately. For the transmission of the encodable packet subsets $\overline{\Q}_p \cap \Pp$ and $\overline{\Q}_s \cap \Ps$, the number of transmissions to ensure that both receivers successfully receive $k_p$ and $k_s$ encodable packets is $B^\txs{a}_3(\mathcal{C})=\max{\{X_1,X_2\}}$. We denote $X_1$ and $X_2$ as the random variables representing the number of transmissions needed to independently deliver $k_p$ packets to the PR and $k_s$ packets to the SR. Note that even though the philosophy of the derivation for $B^\txs{a}_3(\mathcal{C})$ here is the same as $B^\txs{s}_3(\mathcal{C})$, the practical meaning is different. In the ANC scheme, the combinations of the encodable packets are adaptive based on the feedback of both receivers. Thus, the number of transmissions for the coded packets can be represented by the maximum number of transmissions for each receiver requiring its lost packets respectively, with $\mathbb{P}\figbrc{B^\txs{a}_3(\mathcal{C})\leq k} = \mathbb{P}\figbrc{X_1\leq k}\mathbb{P}\figbrc{X_2\leq k}$. We compute the probabilities for arbitrary values of $X_1$ and $X_2$ as shown in \eqref{eq:X1} and \eqref{eq:X2}, which are both negative binomial distributed as
\begin{subequations}
\begin{empheq}[left=\empheqlbrace]{align}
\mathbb{P}\figbrc{X_1=k_p+i} &= \binom{k_p+i-1}{i}p^i_{21}(1-p_{21})^{k_p} \label{eq:X1}\\
\mathbb{P}\figbrc{X_2=k_s+j} &= \binom{k_s+j-1}{j}p^j_2(1-p_2)^{k_s}. \label{eq:X2}
\end{empheq}
\end{subequations}
Both distributions can subsequently be approximated as
\begin{subequations}
\begin{empheq}[left=\empheqlbrace]{align}
&X_1 \sim \mathcal{N}\brc{\mu_{X_1}=\frac{k_p}{1-p_{21}},\; \sigma_{X_1}=\sqrt{\frac{k_pp_{21}}{(1-p_{21})^2}}} \nonumber\\
&X_2 \sim \mathcal{N}\brc{\mu_{X_2}=\frac{k_s}{1-p_2},\; \sigma_{X_2}=\sqrt{\frac{k_sp_2}{(1-p_2)^2}}}. \nonumber
\end{empheq}
\end{subequations}
Therefore, we derive
\begin{IEEEeqnarray}{rCl}
&&\mathbb{P}\figbrc{B^\txs{a}_3(\mathcal{C})=k} = \mathbb{P}\figbrc{B^\txs{a}_3(\mathcal{C})\leq k} - \mathbb{P}\figbrc{B^\txs{a}_3(\mathcal{C})\leq k-1} \nonumber\\
  &=& \mathbb{P}\figbrc{X_2=k}\sum^{k-k_p}_{i=0}\mathbb{P}\figbrc{X_1=k_p+i} + \mathbb{P}\figbrc{X_1=k}\sum^{k-1-k_s}_{j=0}\mathbb{P}\figbrc{X_2=k_s+j}. \label{eq:Pb3ca}
\end{IEEEeqnarray}

Considering again that the distribution of the maximum can be approximated by a normal distribution when the difference between the two standard deviations is small, we approximate $B^\txs{a}_3(\mathcal{C})$ by a normal distribution with mean and standard deviation as
\begin{align}
\mu\brc{B^\txs{a}_3(\mathcal{C})}= &\mu_{X_1}\Phi\brc{\frac{\mu_{X_1}-\mu_{X_2}}{\theta}} + \mu_{X_2}\Phi\brc{\frac{\mu_{X_2}-\mu_{X_1}}{\theta}} + \theta\phi\brc{\frac{\mu_{X_1}-\mu_{X_2}}{\theta}} \nonumber\\
\sigma^2\brc{B^\txs{a}_3(\mathcal{C})}= &\brc{\sigma^2_{X_1}+\mu^2_{X_1}}\Phi\brc{\frac{\mu_{X_1}-\mu_{X_2}}{\theta}} + \brc{\sigma^2_{X_2}+\mu^2_{X_2}}\Phi\brc{\frac{\mu_{X_2}-\mu_{X_1}}{\theta}} \nonumber\\
&+ \brc{\mu_{X_1}+\mu_{X_2}}\theta\phi\brc{\frac{\mu_{X_1}-\mu_{X_2}}{\theta}} - \mu^2\brc{B^\txs{a}_3(\mathcal{C})},
\end{align}
where $\mu_{X_1}$ and $\mu_{X_2}$ are the mean value of the number of the encodable packets in each subset.

As a result, we can approximate the number of retransmissions for each subset in \eqref{eq:Pn3a} by a normal distribution with the moment parameters in \eqref{eq:dist_n3a} below. Subsequently, the expected frame size for the ANC scheme can be determined.
\begin{subequations} \label{eq:dist_n3a}
\begin{eqnarray}
B^\txs{a}_3(\mathcal{C}) &\sim& \mathcal{N}\brc{\mu\brc{B^\txs{a}_3(\mathcal{C})},\; \sigma\brc{B^\txs{a}_3(\mathcal{C})}} \\
B^\txs{a}_3(\overline{\Q}_p \cap \overline{\mathcal{P}}_p) &\sim& \mathcal{N}\brc{\frac{N_pp_1(1-q)p_{12}}{(1-p_1q)(1-p_{21})},\; \sqrt{\frac{N_pp_1(1-q)p_{12}}{1-p_1q}\cdot \frac{p_{21}}{(1-p_{21})^2}}} \\
B^\txs{a}_3(\overline{\Q}_s \cap \overline{\mathcal{P}}_s) &\sim& \mathcal{N}\brc{\frac{N_sp_2p_{21}}{1-p_2},\; \sqrt{N_sp_2p_{21}\cdot \frac{p_2}{(1-p_2)^2}}}
\end{eqnarray}
\end{subequations}

\vspace{2pt}
\subsection{Throughput Improvement of the Network Coding Schemes}
In Subsection \ref{s2-1}, we described the transmission strategies for the conventional ARQ scheme and the two network-coding schemes, and a performance comparison was provided based on an example. The comparison demonstrated that applying network coding can provide performance improvements and the conventional ARQ transmission may provide a lower bound on the system throughput, which is the case that no retransmitted packet is encodable. In this section, we detail a throughput analysis of the improvements of the network coding schemes as compared to the conventional ARQ scheme. From the definition of throughput in Subsection \ref{s2-2} it is clear that the transmission efficiency decreases when more transmission attempts that are needed for delivering a certain fixed number of packets. Therefore, when a given number of packets are transmitted, we only need to show that less transmission attempts are required using the network coding schemes as compared to the ARQ scheme.


In our work, the network coding process is developed in the retransmission session. Since the primary transmission session is the same in all three schemes and so is the secondary transmission session for the two network coding schemes, we mainly analyze the expected number of retransmissions. Starting with the SNC scheme, the expected number of transmission attempts for the retransmission session can be derived by its conditional expectation as shown in \eqref{eq:En3s}. By polynomial expansion, we transform the conditional expectation $E_{\boldsymbol{B}}[B^\txs{s}_3\mid L_p=k_p,L_s=k_s]$ in \eqref{eq:CondEn3s} into
\begin{align*}
&E_{\boldsymbol{B}}[B^\txs{s}_3\mid L_p=k_p,L_s=k_s] \\
= &\frac{k_p}{1-p_{21}} + \frac{k_s}{1-p_2} - \sum^{k_p}_{i=0}\sum^{k_s}_{j=0}\binom{k_p}{i}(1-p_{12})^ip^{k_p-i}_{12} \cdot \binom{k_s}{j}(1-p_{21})^jp^{k_s-j}_{21} \cdot \frac{k_{\min}}{1-p_2p_{21}},
\end{align*}
and obviously in the worst case when there are no encodable packets, $E_{\boldsymbol{B}}[B^\txs{s}_3\mid L_p=k_p,L_s=k_s] = \frac{k_p}{1-p_{21}}+\frac{k_s}{1-p_2}$. Therefore, the upper bound of $E_{\boldsymbol{B}}[B^\txs{s}_{3}]$ becomes
\begin{align*}
E_{\boldsymbol{B}}[B^\txs{s}_3] \leq &\sum^{N_p}_{k_p=0}\sum^{N_s}_{k_s=0}{\mathbb{P}\figbrc{L_p=k_p,L_s=k_s} \cdot \brc{\frac{k_p}{1-p_{21}}+\frac{k_s}{1-p_2}}} \\
= &\frac{N_p p_1(1-q)}{(1-p_1q)(1-p_{21})} + \frac{N_s p_2}{1-p_2}.
\end{align*}
Together with the expected number of transmissions in the first two transmission sessions,
\begin{equation}
\overline{B}^\txs{s}_{\text{afs}} \leq \frac{N_p}{1-p_1q} + \frac{N_p p_1(1-q)}{(1-p_1q)(1-p_{21})} + \frac{N_s}{1-p_2}. \label{eq:app}
\end{equation}
We observe that the result is exactly the same as the expected frame size of the ARQ scheme shown in \eqref{eq:afs_c}, which reflects the worst case that all the retransmitted packets need to be transmitted separately.

Likewise for the ANC scheme, the expected number of transmission attempts for the retransmission session is shown in \eqref{eq:En3a}. The infinite summation $\sum^{\infty}_{k=\max\{k_p,k_s\}}{k \mathbb{P}\figbrc{B^\txs{a}_3(\mathcal{C})=k}}$ in \eqref{eq:CondEn3a} can be bounded by $\frac{k_p}{1-p_{21}} + \frac{k_s}{1-p_2}$ (see \eqref{eq:A1} for a detailed derivation). Thus we can simplify the conditional expectation for $B^\txs{a}_3$ as
\begin{align*}
&E_{\boldsymbol{B}}[B^\txs{a}_3\mid L_p=k_p,L_s=k_s] \leq \frac{k_p}{(1-p_{21})(1-p_{12})} + \frac{k_s}{(1-p_{21})(1-p_2)}.
\end{align*}
Plugging this result back into \eqref{eq:En3a}, we derive the upper bound of $E[B^\txs{a}_3]$ as (see \eqref{eq:A2})
\begin{equation}
E[B^\txs{a}_3] \leq \frac{N_p p_1(1-q)}{(1-p_1q)(1-p_{21})} + \frac{N_s p_2}{1-p_2}, \label{eq:appEn3a}
\end{equation}
and accordingly $\overline{B}^\txs{a}_{\text{afs}}$, which is the same as the SNC scheme. As a result the throughput performance of the ARQ scheme is a lower bound for all three schemes. However, the worst-case scenario for the network coding schemes is when the encodable packet set is empty and the secondary transmitter needs to retransmit the primary and secondary packets individually. As there is a low probability for this to happen, the systems applying the SNC scheme or the ANC scheme generally require a smaller average frame size than the ARQ scheme for delivering the same amount of packets. As a consequence the network coding schemes almost always offer some gain in terms of system throughput.

Furthermore, we provide an analytical proof showing that the ANC scheme almost always outperforms the SNC scheme. Intuitively, the throughput improvement has been demonstrated by example in Subsection \ref{s2-1}. However, an analytical proof is not straightforward since there is no closed-form result for the partial summations of the power series with binomial coefficients in \eqref{eq:A1} for the ANC scheme. As we observed, the only difference for the two network coding schemes lies in the retransmission session; in addition, it is the combination pattern of the encodable packet sets of each scheme which differs. Therefore, we apply induction to determine the expected number of transmissions of the encodable packet sets to imply that this deduction holds for all cases.

Based on the previous two subsets, after the first two transmission sessions, the encodable packet set for each receiver in the retransmission session is $\overline{\Q}_p \cap \Pp$ and $\overline{\Q}_s \cap \Ps$, which is the same for both schemes. We define $L_p=|\overline{\Q}_p \cap \Pp|$ and $L_s=|\overline{\Q}_s \cap \Ps|$ as the number of packets in each encodable packet set, the probability of $k_p$ encodable packets at the PR and $k_s$ at the SR is given by \eqref{eq:Pka}. The performance improvement of the ANC scheme over the SNC scheme is established by showing that the expected number of transmissions of the encodable packets sets for ANC scheme is no larger than the one for SNC scheme. A detailed proof of the mathematical induction method is given in Appendix \ref{App2}. In summary, network coding can provide performance improvement over the conventional ARQ scheme; likewise, the ANC scheme performs at least as well as the SNC, and in most cases, even better, as numerical results show.


\begin{figure*}
\centering
\hspace{-0.35cm}
\subfigure[]{
\includegraphics[width=0.37\textwidth,height=5.7cm]{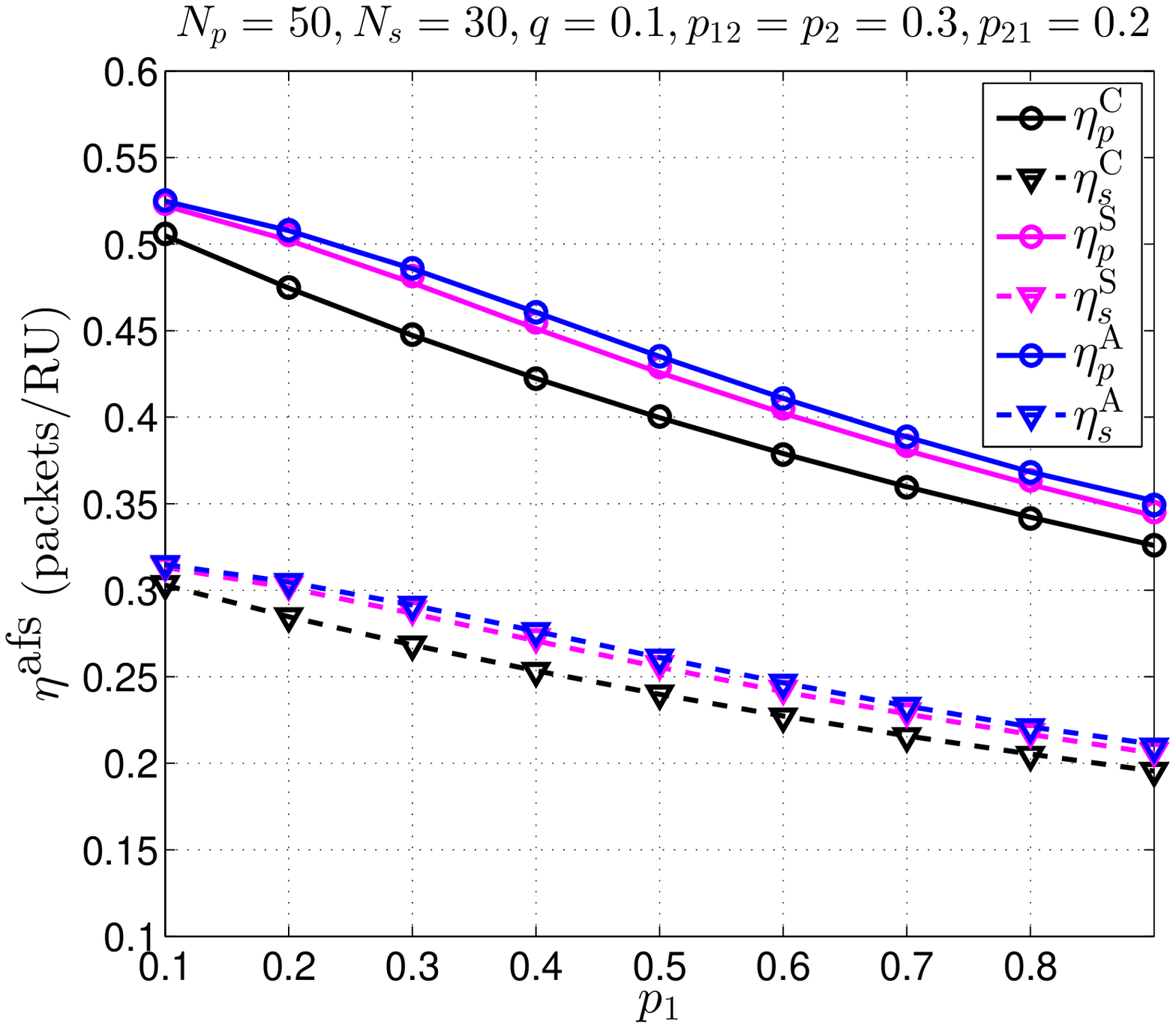}
\label{fig:p1com}} \hfil \hspace{-0.45cm}
\subfigure[]{
\includegraphics[width=0.37\textwidth,height=5.7cm]{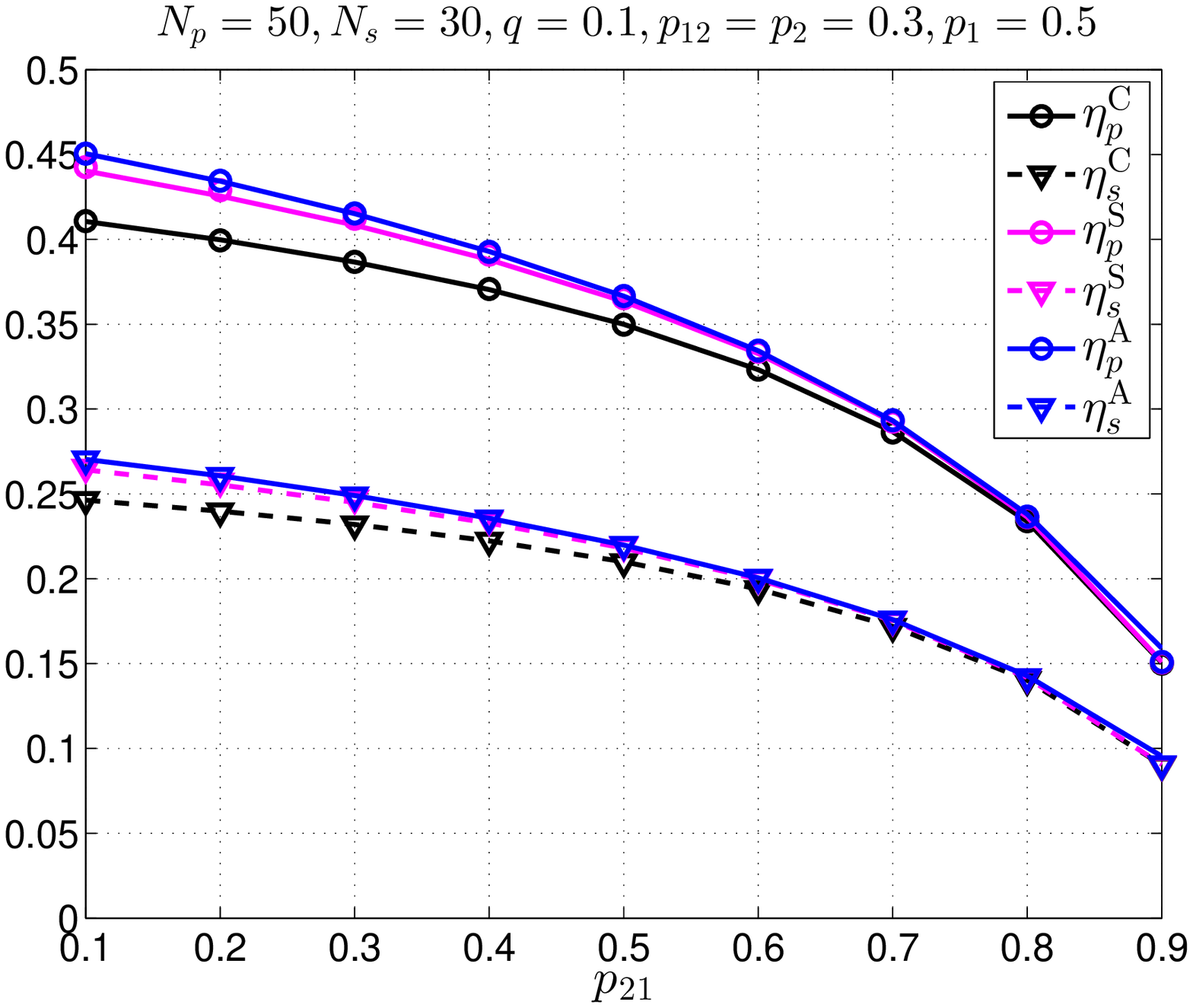}
\label{fig:p21com}}
\caption{Throughput comparison as a function of link qualities $p_1$ and $p_{21}$.}
\label{fig:pcom}
\end{figure*}

We now provide some numerical results in Fig. \ref{fig:pcom} to support our analysis. We compare the performance of the two network coding schemes to the conventional ARQ scheme as a function of the link qualities related to the PR. Theoretical results are shown by lines, while numerical results are indicated by markers. We observe a perfect match, thus validating our derivation above. The performance of the primary system is shown by a solid line, and the secondary system by a dashed line. Black curves denote the conventional ARQ scheme, magenta curves denote the SNC scheme and blue curves denote the ANC scheme. In Fig. \ref{fig:p1com} we compare the packet throughput performance of the three schemes as the direct primary link varies. Both the SNC and the ANC schemes perform better than the ARQ scheme, and the ANC scheme performs better than the SNC scheme. The improvements are insensitive to variations in $p_1$, implying that network coding is effective. Fig. \ref{fig:p21com} shows the performance comparison as a function of the cross link quality from the ST to the PR. When $p_{21}$ varies, the performance of the primary system is degraded vastly from $0.45$ to $0.15$ as the direct link from the PT to the PR is poor. Moreover, as the cross link gets worse, the gain from network coding is vanishing. The comparison gives us an indication that the performance improvement depends on the existence of coding opportunities, which themselves depend on the link qualities.

\vspace{2pt}
\subsection{Accuracy of the Normal Approximation}
\begin{figure}[htp]
\centering
\includegraphics[width=0.44\textwidth]{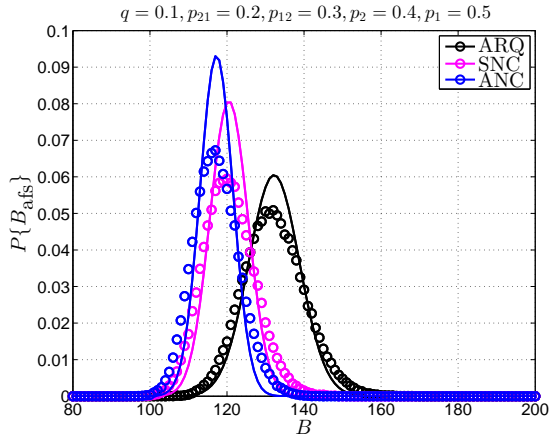}
\caption{Probability mass function of $B_{\text{afs}}$.}
\label{fig:n_pdf}
\end{figure}

To determine the expected frame size for the three transmission schemes in a closed-form expression, we applied the normal approximation detailed in Lemma \ref{Lemma1}. To show the accuracy of the normal approximation to the original distribution, Fig. \ref{fig:n_pdf} compares the experimental results of the total number of transmissions for the three schemes to the approximations when $N_p=50$ and $N_s=30$. The packet erasure probabilities are chosen randomly. Circles denote the experimental results while lines denote the corresponding normal approximations with the same mean value. It is shown that a normal distribution can approximate the distribution of $B$ fairly well. With larger $N_p$ and $N_s$, this approximation performs better which we did not show in this figure.

\section{Performance Analysis for Truncated Frame Size}\label{sec:performance_p}


In Section \ref{sec:performance_a}, we considered the case where the frame size is allowed to grow infinitely large, thus providing a benchmark for lossless transmission. For the adaptive frame-size case the instantaneous frame size is limited to $(N_1+N_2) \leq B \leq \infty$. In this case the distribution of $B$ is well approximated by a normal distribution, $B\sim \mathcal{N}\brc{\mu, \sigma}$, allowing for frame sizes in the interval $-\infty \leq B \leq \infty$. As long as the sum $(N_p+N_s)$ is sufficiently large, the probability of frame lengths smaller than $(N_p+N_s)$ is negligible, as illustrated in Fig. \ref{fig:n_pdf} for $N_p=50$ and $N_s=30$. The normal approximation is therefore useful for analysis and design.

In order to limit reception delays due to arbitrarily large frame sizes, we now truncate the instantaneous frame size to be in the interval $(N_p+N_s) \leq B \leq \widehat{B}$, with a maximum frame size of $\widehat{B}$. However, as discussed in Subsection \ref{s2-2}, when enforcing such a constraint the resulting scheme is no longer lossless. There is now a non-zero probability of outage, $P_{\text{out}}(\widehat{B}))=\mathbb{P}\{B> \widehat{B}\}$, that some primary and/or secondary packets may not be delivered successfully within a frame. The truncated scheme is therefore characterized by an averaged throughput under the constraint of a given acceptable outage probability.

For the analysis of the truncated frame-size case, we follow a similar approach as for the adaptive frame size. Applying the same reasoning we can again neglect the lower limit on the frame size, and thus consider the frame-size interval $-\infty \leq B \leq \widehat{B}$ instead. However, due to the upper-truncation, we now approximate the distribution of $B$ by an upper truncated normal distribution, $B\sim \mathcal{TN}\brc{\hat{\mu},\hat{\sigma}}$, simply by truncating the approximating normal distribution for the adaptive case.

Let $\phi(0,1;x)$ denote the pdf of a standard normal distribution with argument $x$, and  let $\Phi(0,1;x)$ denote the cdf of a standard normal distribution with argument $x$, respectively. Following the definitions of truncated normal distributions in \cite{JB14,CUD94}, the mean and the variance of the upper-truncated normal distribution are
\begin{align}
&\hat{\mu} = \mu - \sigma \cdot \frac{\phi(0,1;\beta)}{\Phi(0,1;\beta)},
\hspace{10mm}\hat{\sigma}^2 = \sigma^2 \cdot \brc{1 - \frac{\beta\phi(0,1;\beta)}{\Phi(0,1;\beta)} - \brc{\frac{\phi(0,1;\beta)}{\Phi(0,1;\beta)}}^2}.
\end{align}
respectively. Here $\beta = (\widehat{B}-\mu)/{\sigma}$, 
while $\mu$ and $\sigma$ are the mean and variance of the general normal distribution $\mathcal{N}\brc{\mu, \sigma}$. Then, formally the upper truncated normal pdf and cdf can be evaluated by the general normal distribution as:
\begin{align}
&\psi(\hat{\mu},\hat{\sigma};B)=
\begin{cases}
  \frac{\phi(\mu,\sigma;B)}{\Phi(\mu,\sigma;\widehat{B})}     & \text{if } B \leq \widehat{B} \\
  0                                                           & \text{if } B > \widehat{B}
\end{cases}, 
&\Psi(\hat{\mu},\hat{\sigma};B)=
\begin{cases}
  \frac{\Phi(\mu,\sigma;B)}{\Phi(\mu,\sigma;\widehat{B})}     & \text{if } B \leq \widehat{B} \\
  1                                                           & \text{if } B > \widehat{B}
\end{cases}.\label{eq:tn_pdf-cdf}
\end{align}
For each of the three transmission schemes, the transmission process can be approximated by an upper truncated normal distribution as $B\sim \mathcal{TN}\brc{\hat{\mu},\hat{\sigma}}$.

\vspace{2pt}
\subsection{Throughput-delay Tradeoff Analysis}
In a truncated system with fixed frame size $\widehat{B}$, the throughput-delay tradeoff can be balanced by an upfront evaluation of the packet transmission scheme.
If the outage probability is controlled within a certain range, a corresponding packet throughput can be achieved by estimating the number of packets to be transmitted in the following frame. At the beginning of each frame a pair of $(N_p,N_s)$ is estimated, based on the averaged behaviour detemined by the outage probability. Given a value $0\leq P_{\text{out}}(\widehat{B})\leq 1$, we seek $B\leq \widehat{B}$ satisfying:
\begin{align*}
&P_{\text{out}}(\widehat{B}) = 1-\psi(\mu,\sigma;\widehat{B}) = \mathrm{Q}\brc{\frac{\widehat{B}-\mu}{\sigma}}.
\end{align*}
Note that the cdf of the general normal distribution can be represented by a Q-function, and thus $\widehat{B}$ can be represented by the inverse Q-function as $\widehat{B} = \mu + \sigma\cdot \mathrm{Q}^{-1}(P_{\text{out}})$.
With the frame size $\widehat{B}$ fixed, the mean and variance of the approximated general normal distribution $\mathcal{N}\brc{\mu,\sigma}$ can be derived. 

In the truncated frame-size case, there is no transmission when $B>\widehat{B}$. As a consequence, the approximation of the transmission process needs to be adjusted to the upper truncated normal distribution with $\mathbb{P}\{B\;|\;B> \widehat{B}\} = 0$. Based on the general normal distribution $\mathcal{N}(\mu,\sigma)$, the approximated $\mathcal{TN}\brc{\hat{\mu},\hat{\sigma};\widehat{B}}$ can be determined by \eqref{eq:tn_pdf-cdf}.
Accordingly, the adjusted number of packets $(\hat{N}_p,\hat{N}_s)$ to be transmitted is obtained. Arranging the number of packets to be transmitted at the beginning of each frame properly can reduce the risk of large queueing delays. 

\vspace{2pt}
\subsection{Accuracy of the Upper Truncated Normal Approximation}

\begin{figure*}[tp]
\centering
\subfigure[$N_s = 19$]{ \hspace{-5mm}
\includegraphics[width=0.34\textwidth,height=4.9cm]{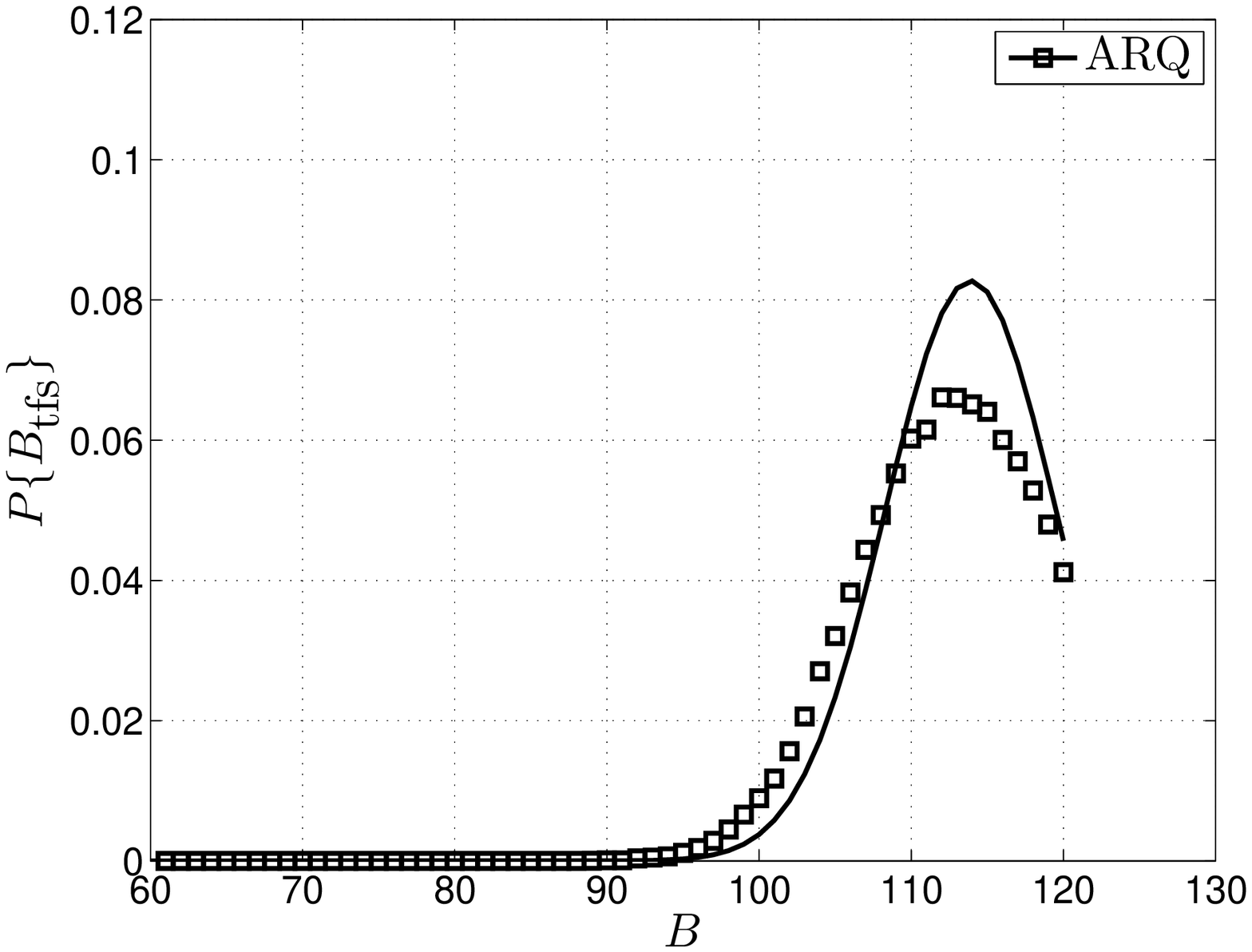}
}\hfil
\subfigure[$N_s = 25$]{ \hspace{-3mm}
\includegraphics[width=0.31\textwidth,height=4.9cm]{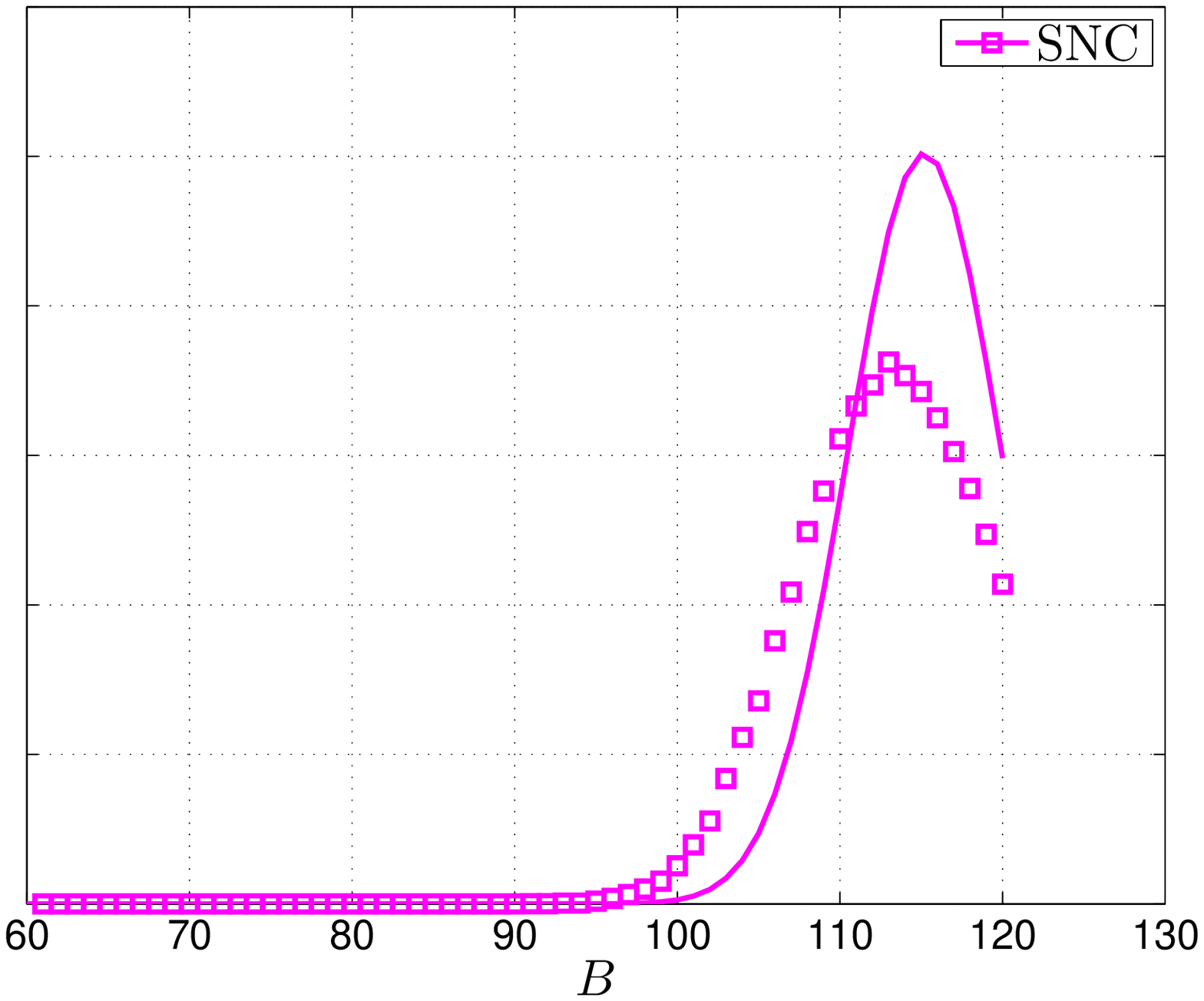}
}\hfil
\subfigure[$N_s = 27$]{ \hspace{-3mm}
\includegraphics[width=0.31\textwidth,height=4.9cm]{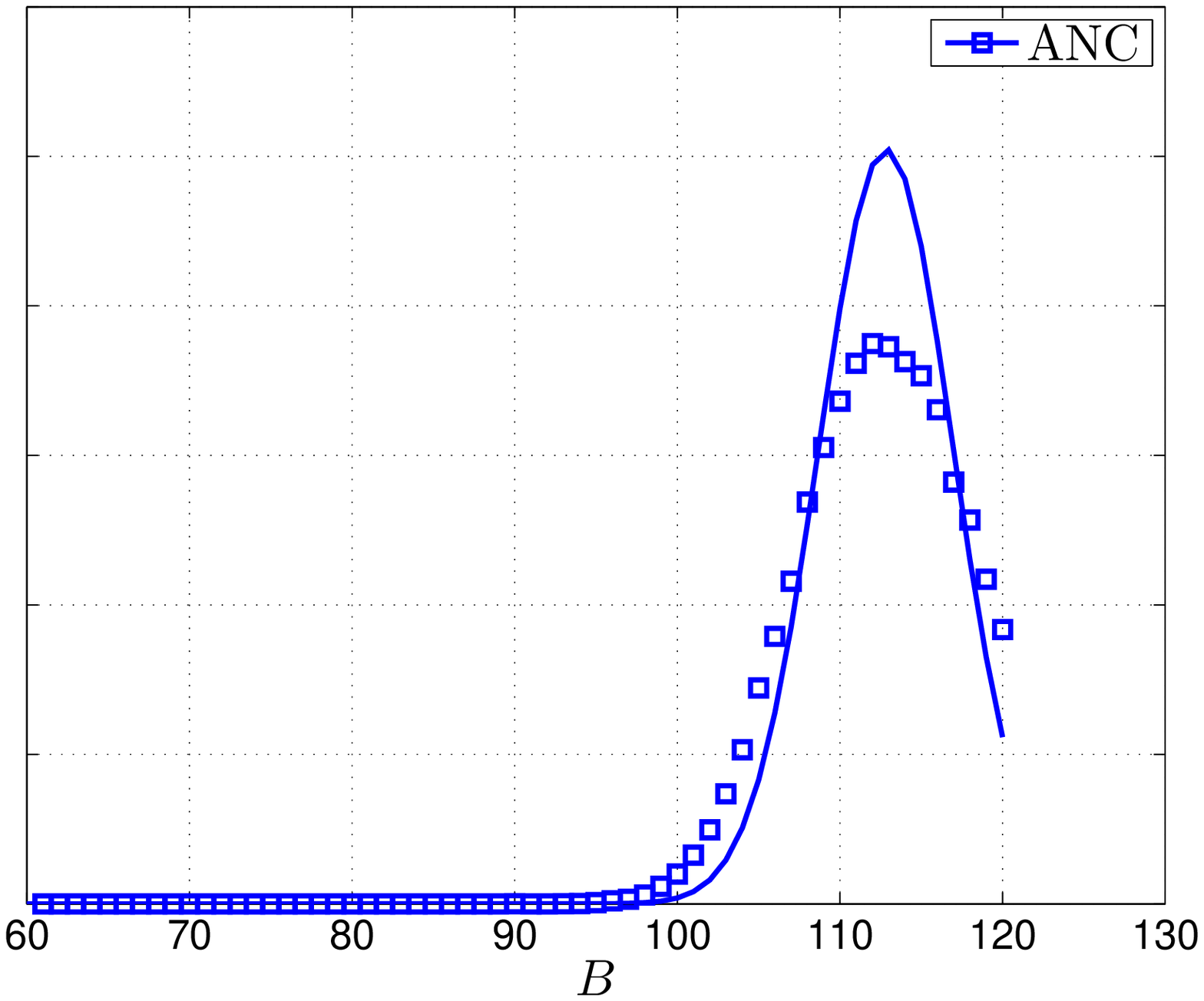}
}
\caption{Probability mass function of $B_{\text{tfs}}$.}
\label{fig:tn_pdf}
\end{figure*}

In the same simulation environment as the adaptive frame-size case in Fig. \ref{fig:n_pdf}, the experimental results of the total number of transmissions for the three schemes are compared to the truncated normal approximations when $P_{\text{out}}(\widehat{B})= 0.1$ in the general normal approximation in Fig. \ref{fig:tn_pdf}. Squares denote the experimental results while lines denote the corresponding truncated normal approximations. The accuracy of the upper truncated normal approximation is shown by an example where the frame size is set as $\widehat{B}=120$ and the number of primary packets to be transmitted is predefined as $N_p=50$. In this case, the number of secondary packets which can be delivered in each frame determines the throughput performance of each scheme. To satisfy the requirement of the outage probability, the experimental results show that $19$ secondary packets can be delivered successfully when applying the ARQ scheme, while $25$ secondary packets can be delivered when applying the SNC scheme and $27$ secondary packets delivered with the ANC scheme.


%
%
\section{Numerical Results}\label{sec:numresults}

To keep consistency of the numerical experiments, the simulation environment is a stationary network of packet erasure links with erasure probabilities $q=0.1$, $p_{21}=0.2$, $p_{12}=0.3$, $p_2=0.4$ and $p_1=0.5$. We first consider the impact of the number of packets to be delivered on throughput performance in the adaptive frame-size case. After that we investigate the throughput performance with varying frame size in the truncated frame-size case.

\begin{figure*}[t]
\centering
\subfigure[$N_p = 50$ (afs)]{
\includegraphics[width=0.37\textwidth,height=5.7cm]{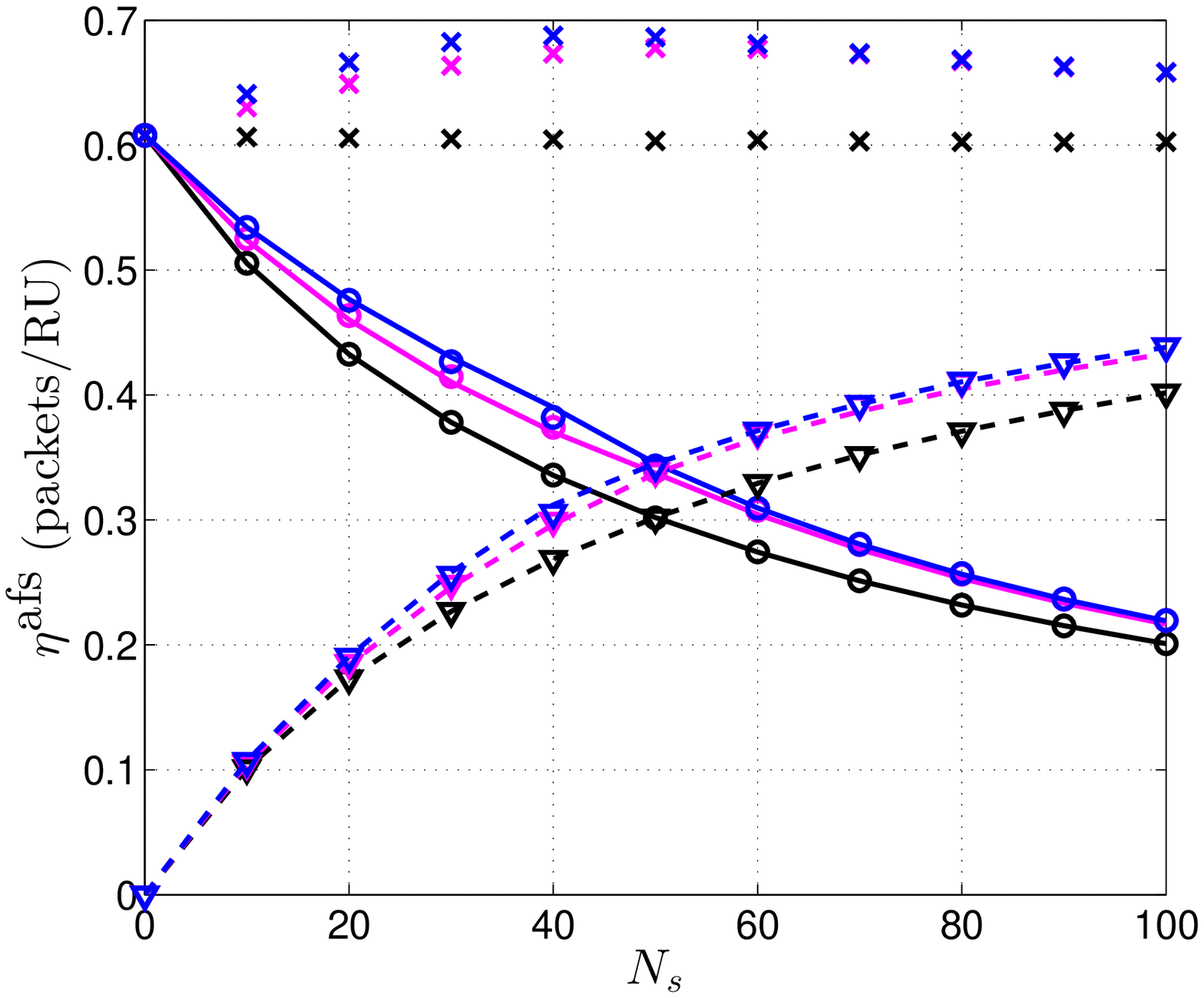}
\label{fig:flexn2com}} \hfil
\subfigure[$N_s = 50$ (afs)]{
\includegraphics[width=0.395\textwidth,height=5.7cm]{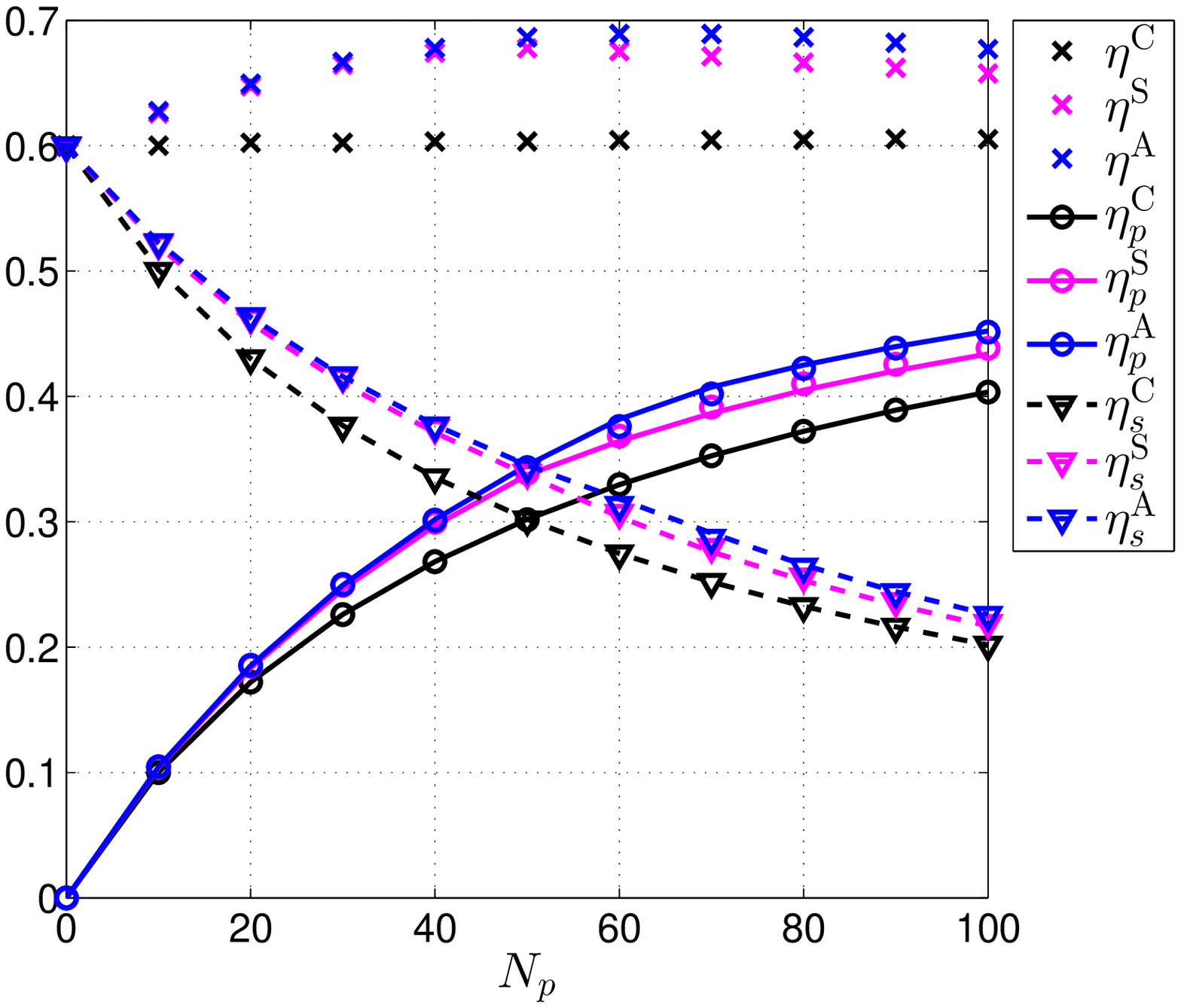}
\label{fig:flexn1com}}
\bigskip
\subfigure[$N_p = 50$ (tfs)]{
\includegraphics[width=0.37\textwidth,height=5.7cm]{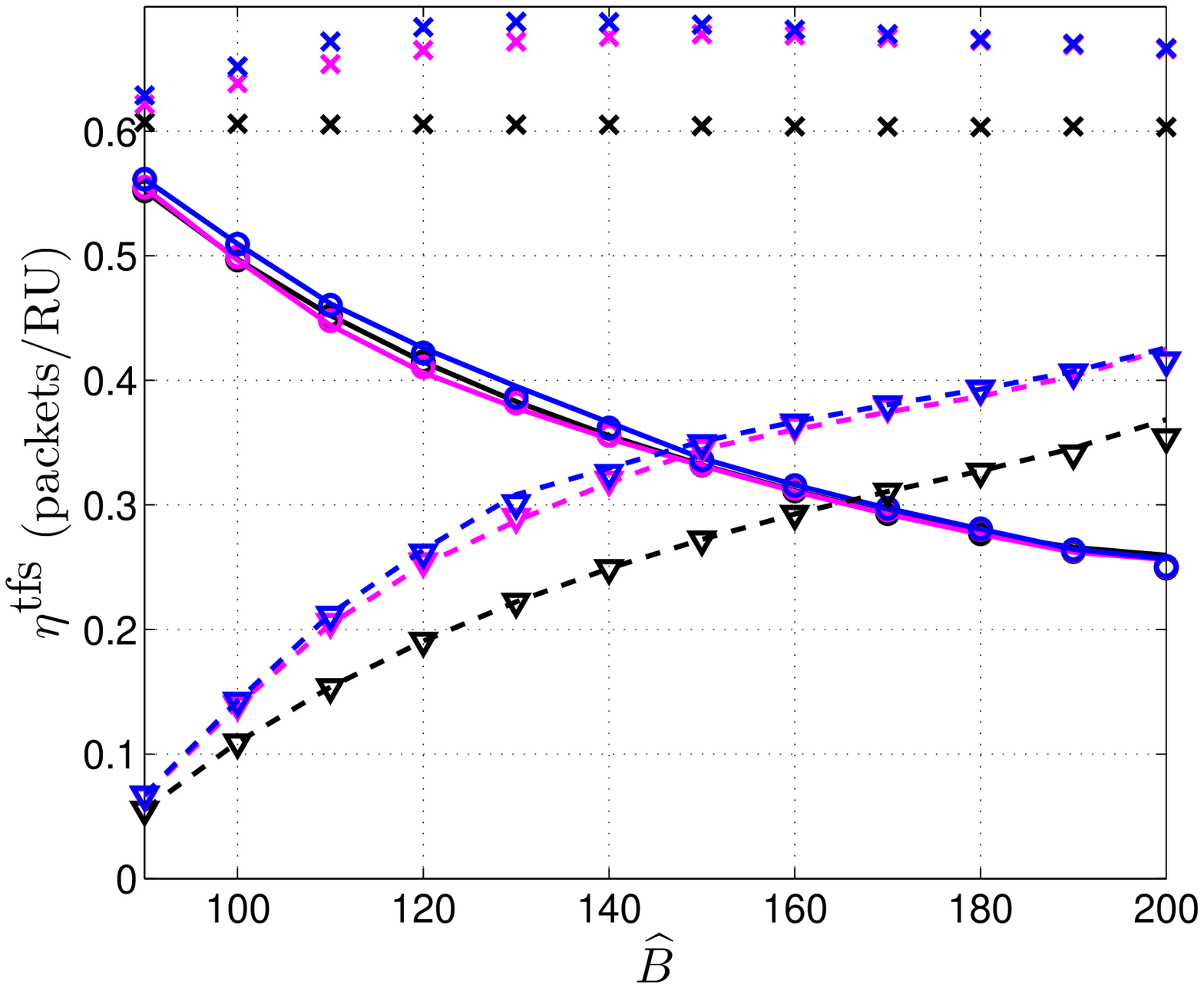}
\label{fig:fixn2com}} \hfil
\subfigure[$N_s = 50$ (tfs)]{
\includegraphics[width=0.395\textwidth,height=5.7cm]{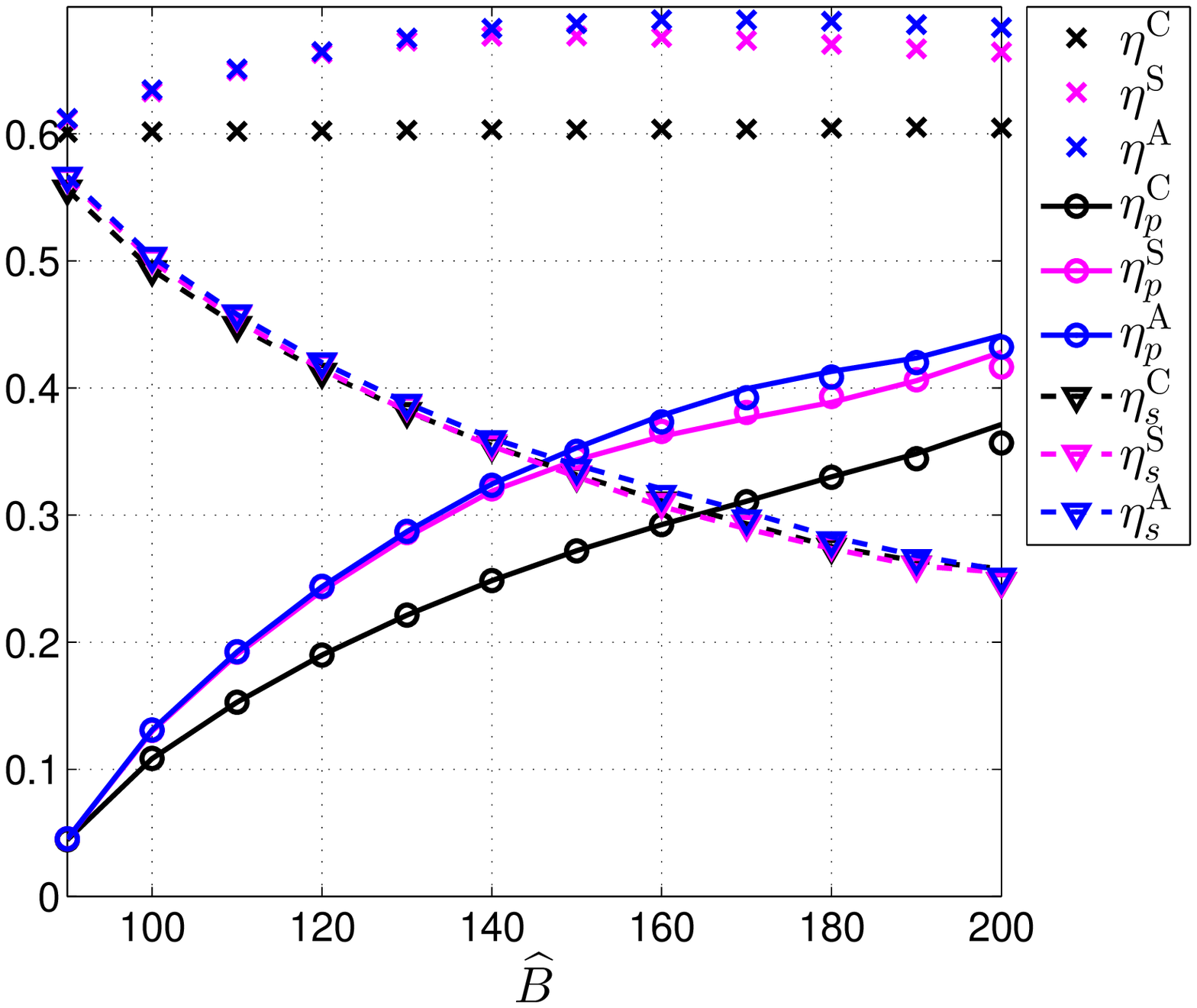}
\label{fig:fixn1com}}
\caption{Throughput comparison as a function of $N_p$ and $N_s$.}
\label{fig:N_com}
\end{figure*}

Since the packet throughput is defined as the packet transmission efficiency, then for each pair of $(N_p,N_s)$, there is a pair of corresponding $(\eta_p,\eta_s)$ which indicates the system performance. In the adaptive frame-size case, we fix $N_p=50$ to show the throughput variation as a function of $N_s$, and vice versa to show the throughput performance for $N_s=50$ as a function of $N_p$. We also show the accuracy of the normal approximation by comparing the experimental results and the approximations. The circles denote the experimental results for the throughput of the primary system, the triangles denote the experimental results for the throughput of the secondary system. Meanwhile, the solid lines denote the normal approximation for the primary system and the dashed lines denote the normal approximation for the secondary system. The overall throughput of the network is denoted by the stars. Obviously, the normal approximation we applied matches the experimental results quite well. Fig. \ref{fig:flexn2com} provides the throughput comparison for the three transmission schemes when $N_p$ is fixed as $50$. With increasing $N_s$, the secondary throughput increases while the primary throughput decreases. Similarly in Fig. \ref{fig:flexn1com}, an increasing $N_p$ leads to increasing primary throughput and decreasing secondary throughput. In both cases, the use of network coding provides an improvement in performance for both primary and secondary systems and the ANC scheme outperforms the SNC scheme.

In the truncated frame-size case, we fix $N_p=50$ in Fig. \ref{fig:fixn2com} and $N_s=50$ in Fig. \ref{fig:fixn1com} to show the throughput variation as a function of $\widehat{B}$. The results show that the upper truncated normal distribution can properly approximate the experimental transmissions in both scenarios. When the number of primary packets $N_p$ is fixed, the throughput variation is mainly determined by the secondary packet transmissions as the size of each frame is predefined. Thus the throughput of the secondary system increases with increasing $\widehat{B}$. Besides, there is little difference among the primary throughput of the three schemes when both $N_p$ and $\widehat{B}$ are fixed. 

Furthermore, there are some interesting observations of the overall throughput performance in both cases. We see that both network coding schemes perform better than the ARQ scheme, whilst the ANC scheme outperforms the SNC scheme especially when $N_p\geq N_s$. This is because in the experimental environment we assume the link between the ST and the PR has a better quality than the link between the ST and the SR, i.e., $p_{21}\leq p_2$. It reflects that the link quality affects the transmissions with different schemes applied, as we compared in Fig. \ref{fig:pcom}, and it is an important factor to consider when making decisions on resource sharing and collaboration between the primary and secondary system. In addition, the overall throughput keeps consistent when applying the ARQ scheme while it achieves an optimum at some point when applying the network coding schemes. It indicates that an appropriate management of cooperative communication between the primary system and the secondary system can lead to a better performance when network coding is applied.


%
%
\section{Conclusions}\label{sec:conclusion}

We have investigated the impact of cooperation to gain more transmission opportunities for the secondary system in cognitive radio networks. By relaying the primary message during the retransmission phase, the secondary transmitter obtains opportunities for transmission. Compared to a conventional ARQ transmission scheme, we developed two network coding schemes in which the secondary transmitter cooperates by conducting the retransmission sessions for both the primary and the secondary systems. 

We first divide the transmission process into three transmission sessions for the three transmission schemes, and then subsequently analyze each of the sessions. The performance of each scheme was investigated analytically for two cases, the adaptive frame-size case and the truncated frame-size case. In the adaptive frame-size case, the system throughput is measured by the total expected number of transmission attempts and the system is lossless; in the truncated frame-size case, both the throughput and the outage probability are considered where the system is defined as in outage when there exists packet loss. For simplicity of analysis, we approximated the distribution of the number of transmission attempts in both cases. For the case of adaptive frame size, a general normal approximation was proposed, based on which, a truncated normal approximation is further generated for the case of a truncated frame size. We also compared the system throughput based on experimental results to the approximations. The results show that a normal distribution can approximate the transmission process well and it can reduce the complexity of computations. 



%
%
\appendix
\subsection{Proof of Transmission Efficiency of A Two-receiver Broadcast Channel} \label{App1}
\renewcommand{\theequation}{\thesection.\arabic{equation}}

\begin{IEEEproof}
For a two-receiver broadcast channel using typical ARQ scheme, the receiver immediately sends a Nack when there is a packet loss and this packet has not been received successfully before. In our system of retransmission session, the PR and the SR are with packet erasure probabilities of $p_{21}$ and $p_2$ from the ST. Let $X_1$ and $X_2$ be the random variables denoting the numbers of transmission attempts to successfully deliver a packet to the PR and the SR, respectively. The number of transmissions to ensure that both receivers successfully receive this packet is the random variable $Y=\max\{X_1,X_2\}$ with
\begin{align}
\mathbb{P}\figbrc{Y\leq k} &= \mathbb{P}\figbrc{X_1\leq k}\mathbb{P}\figbrc{X_2\leq k} \nonumber\\
&= \sum^{k}_{i=1}p^{i-1}_{21}(1-p_{21}) \sum^{k}_{j=1}p^{j-1}_{2}(1-p_{2}) \nonumber\\
&= \brc{1-p^k_{21}}\brc{1-p^k_{2}}.
\end{align}
Therefore
\begin{align}
\mathbb{P}\figbrc{Y=k} &= \mathbb{P}\figbrc{Y\leq k} - \mathbb{P}\figbrc{Y\leq k-1} \nonumber\\
&= \brc{1-p^k_{21}}\brc{1-p^k_{2}} - \brc{1-p^{k-1}_{21}}\brc{1-p^{k-1}_{2}}
\end{align}
and the average number of transmission attempts per packet is
\begin{align}
\mu^\textsc{bc}(2) &= E[Y] = \sum^{\infty}_{k=1} k \mathbb{P}\figbrc{Y=k} \nonumber\\
&= \sum^{\infty}_{k=1} k \brc{p^{k-1}_{21}\brc{1-p_{21}}+p^{k-1}_2\brc{1-p_2}-p^{k-1}_{21}p^{k-1}_2\brc{1-p_{21}p_2}} \nonumber\\
&= \frac{1}{1-p_{21}} + \frac{1}{1-p_2} - \frac{1}{1-p_2p_{21}},
\end{align}
in which we let the first power series $\sum^{\infty}_{k=1} k p^{k-1}_{21}$ denoted by $S$ and it can be transformed as
\begin{align}
S = \sum^{\infty}_{k=0} (k+1) p^k_{21} &= \sum^{\infty}_{k=0} k p^k_{21} + \sum^{\infty}_{k=0} p^k_{21} = S\cdot p_{21}+\frac{1}{1-p_{21}}. \nonumber
\end{align}
Eventually we get $S = \frac{1}{\brc{1-p_{21}}^2}$ and the close-form result of the left two power series accordingly.
\end{IEEEproof}


\subsection{Proof of Throughput Improvement of Network Coding} \label{App2}

\renewcommand{\theequation}{\thesection.\arabic{equation}}

\begin{IEEEproof}
\begin{figure*}
\normalsize
\begin{eqnarray}
\hspace{-10.2pt}
&& \sum^{\infty}_{k=\max\{k_p,k_s\}}{k \mathbb{P}\figbrc{B^\txs{a}_3(\mathcal{C})=k}} \label{eq:A1} \\
  &=& \sum^{\infty}_{k}{k \mathbb{P}\figbrc{X_2=k}\sum^{k-k_p}_{i=0}\mathbb{P}\figbrc{X_1=k_p+i}} + \sum^{\infty}_{k}{k \mathbb{P}\figbrc{X_1=k}\sum^{k-1-k_s}_{j=0}\mathbb{P}\figbrc{X_2=k_s+j}} \nonumber\\
  &\stackrel{(a)}=& \sum^{\infty}_{k}{k \mathbb{P}\figbrc{X_2=k}\sum^{\infty}_{i=0}\mathbb{P}\figbrc{X_1=k_p+i}} + \sum^{\infty}_{k}{k \mathbb{P}\figbrc{X_1=k}\sum^{\infty}_{j=0}\mathbb{P}\figbrc{X_2=k_s+j}} \nonumber\\
  &&-\sum^{\infty}_{k}{k \mathbb{P}\figbrc{X_2=k}\sum^{\infty}_{i=k-k_p+1}\mathbb{P}\figbrc{X_1=k_p+i}} - \sum^{\infty}_{k}{k \mathbb{P}\figbrc{X_1=k}\sum^{\infty}_{j=k-k_s}\mathbb{P}\figbrc{X_2=k_s+j}} \nonumber\\
  &\stackrel{(b)}\leq& \sum^{\infty}_{k}{k \mathbb{P}\figbrc{X_2=k}} + \sum^{\infty}_{k}{k \mathbb{P}\figbrc{X_1=k}} \nonumber\\
  &\stackrel{(c)}\leq& \frac{k_p}{1-p_{21}} + \frac{k_s}{1-p_2} \nonumber
\end{eqnarray}
\end{figure*}
\begin{figure*}
\normalsize
\begin{eqnarray}
\hspace{-8.2pt}
E[B^\txs{a}_3] &=& \sum^{N_p}_{k_p=0}\sum^{N_s}_{k_s=0}{\mathbb{P}\figbrc{L_p=k_p,L_s=k_s}E[B^\txs{a}_3\mid L_p=k_p,L_s=k_s]} \label{eq:A2} \\
               &\leq& \sum^{N_p}_{k_p=0}\sum^{N_s}_{k_s=0}{\mathbb{P}\figbrc{L_p=k_p,L_s=k_s} \cdot \brc{\frac{k_p}{(1-p_{21})(1-p_{12})} + \frac{k_s}{(1-p_{21})(1-p_2)}}} \nonumber\\
               &=& \frac{N_p p_1(1-q)}{(1-p_1q)(1-p_{21})} + \frac{N_s p_2}{1-p_2} \nonumber
\end{eqnarray}
\hrulefill
\vspace*{4pt}
\end{figure*}
When analyzing the performance improvement of the ANC scheme over the conventional ARQ scheme, we derive the upper bound of the expected number of retransmissions $E[B^\txs{a}_3]$ by simplifying $E_{\boldsymbol{B}}[B^\txs{a}_3\mid L_p=k_p,L_s=k_s]$ shown in \eqref{eq:CondEn3a}, in which the infinity summation can be reduced as shown in \eqref{eq:A1}. With the probability of $B^\txs{a}_3(\mathcal{C})$ derived in \eqref{eq:Pb3ca}, $(a)$ follows from defining the partial sum as the difference of two infinite sums; $(b)$ follows from omitting the last two summations; and $(c)$ applies the fact that the infinite summation starting from $\max\{k_p,k_s\}$ is no larger than one starting from $k_p$ or $k_s$, since there are always $|k_p-k_s|$ items less. Based on this result, the upper bound of $E_{\boldsymbol{B}}[B^\txs{a}_3\mid L_p=k_p,L_s=k_s]$ is obtained straightforwardly. 
Moreover, in \eqref{eq:A2} a detailed derivation of the upper bound of $E_{\boldsymbol{B}}[B^\txs{a}_3]$ in \eqref{eq:appEn3a} is provided. By substituting the upper bound we derived above, the terms on $k_p$ and $k_s$ can be divided and polynomial expansion is then applied. 


Furthermore, when analyzing the performance improvement of the ANC scheme over the SNC scheme, mathematical induction method is applied in proof as the only difference of the two schemes lies in the transmission of the encodable packets. We only need to show that the expected number of transmissions of the encodable packets for the ANC scheme is no larger than for the SNC scheme to indicate the performance improvement. As we defined $L_p$ and $L_s$ as the number of packets in each encodable packet set, the probability of $k_p$ encodable packets at the PR and $k_s$ at the SR is given by \eqref{eq:Pka}, which is the same for both schemes. Thus the expected number of transmissions of the encodable packets $E[B^\txs{s,a}_3(\mathcal{C})]$ is determined by $E_{\boldsymbol{B}}[B^\txs{s,a}_3(\mathcal{C})\mid L_p=k_p,L_s=k_s]$, which is analyzed in the following cases.
\begin{enumerate}
\item When $k_p=1,k_s=1$: in this case, there is only one encoded packet to be transmitted, thus for either the SNC scheme or the ANC scheme, $E_{\boldsymbol{B}}[B^\txs{s,a}_3(\mathcal{C})\mid L_p=k_p,L_s=k_s]= 1\cdot \mu^\textsc{bc}(2)$ where we had $\mu^\textsc{bc}(2) = \frac{1}{1-p_{21}} + \frac{1}{1-p_2} - \frac{1}{1-p_2p_{21}}$ (See Appendix \ref{App1}). The deduction holds.
\item When $k_p=1,k_s=2$: in this case, for the SNC scheme, there is one encoded packet and one secondary packet to be transmitted separately, thus
    \begin{align*}
    E_{\boldsymbol{B}}[B^\txs{s}_3(\mathcal{C})\mid L_p=k_p,L_s=k_s]= 1\cdot \mu^\textsc{bc}(2)+\frac{1}{1-p_2} = \frac{2}{1-p_2}+\frac{p_{21}(1-p_2)(1-p_2p_{21})}{(1-p_{21})(1-p_2p_{21})^2}.
    \end{align*}

    For the ANC scheme, we have the probability mass function
    \begin{align*}
    \mathbb{P}\figbrc{B^\txs{a}_3(\mathcal{C})=k} = &\mathbb{P}\figbrc{B^\txs{a}_3(\mathcal{C})\leq k} - \mathbb{P}\figbrc{B^\txs{a}_3(\mathcal{C})\leq k-1} \\
    = &\sum^{k}_{i=1} p^{i-1}_{21}\brc{1-p_{21}}\cdot \sum^{k}_{j=2} \binom{j-1}{1} p^{j-2}_2\brc{1-p_2}^2  \nonumber\\
    &- \sum^{k-1}_{i=1} p^{i-1}_{21}\brc{1-p_{21}}\cdot \sum^{k-1}_{j=2} \binom{j-1}{1} p^{j-2}_2\brc{1-p_2}^2, \nonumber
    \end{align*}
    in which we can transform $\sum^{k}_{j=2} \binom{j-1}{1} p^{j-2}_2\brc{1-p_2}^2$ into $\sum^{k-2}_{j'=0} (j'+1) p^{j'}_2\brc{1-p_2}^2$ and accordingly get a closed-form result for this partial summation. Similarly to the rest of the partial summations, we get the probability function above in a closed-form expression.
    \begin{align*}
    &\mathbb{P}\figbrc{B^\txs{a}_3(\mathcal{C})=k} \\
    = &(1-p_{21})p^{k-1}_{21}\brc{(k-1)p^k_2-kp^{k-1}_2+1} + (1-p_2)^2 p^{k-2}_2(k-1)\brc{1-p^{k-1}_{21}} \nonumber
    \end{align*}
    And then the conditional expected number of transmissions of the encodable packets is derived as
    \begin{align*}
E_{\boldsymbol{B}}[B^\txs{a}_3(\mathcal{C})\mid L_p=k_p,L_s=k_s] = &\sum^{\infty}_{k=2}{k \mathbb{P}\figbrc{B^\txs{a}_3(\mathcal{C})=k}} \\
    = &\sum^{\infty}_{k'=0}{(k'+2) \mathbb{P}\figbrc{B^\txs{a}_3(\mathcal{C})=k'+2}} \nonumber\\
    = &\frac{2}{1-p_2} + \frac{-(p_2p_{21})^3+(p_2p_{21})^2+p_2p^2_{21}-p_{21}}{\brc{1-p_2p_{21}}^3}. \nonumber
    \end{align*}

    Comparing the results for two schemes, we find that $E_{\boldsymbol{B}}[B^\txs{a}_3(\mathcal{C})\mid L_p=k_p,L_s=k_s] \leq E_{\boldsymbol{B}}[B^\txs{s}_3(\mathcal{C})\mid L_p=k_p,L_s=k_s]$. Thus the deduction holds.
\item When $k_p=2,k_s=1$: similar to case 2), the deduction holds in this case.
\item When $k_p=2,k_s=2$: in this case, there are two encoded packets for transmission, using the idea of derivation in the previous cases, the deduction can be shown holding.
\item When $k_p$ and $k_s$ continue increasing, we find that the idea of derivation is similar and the deduction should hold for all the following cases.
\end{enumerate}

\end{IEEEproof}


\bibliographystyle{IEEEtran}
\bibliography{journal}

\begin{thebibliography}{10}
\providecommand{\url}[1]{#1}
\csname url@samestyle\endcsname
\providecommand{\newblock}{\relax}
\providecommand{\bibinfo}[2]{#2}
\providecommand{\BIBentrySTDinterwordspacing}{\spaceskip=0pt\relax}
\providecommand{\BIBentryALTinterwordstretchfactor}{4}
\providecommand{\BIBentryALTinterwordspacing}{\spaceskip=\fontdimen2\font plus
\BIBentryALTinterwordstretchfactor\fontdimen3\font minus
  \fontdimen4\font\relax}
\providecommand{\BIBforeignlanguage}[2]{{%
\expandafter\ifx\csname l@#1\endcsname\relax
\typeout{** WARNING: IEEEtran.bst: No hyphenation pattern has been}%
\typeout{** loaded for the language `#1'. Using the pattern for}%
\typeout{** the default language instead.}%
\else
\language=\csname l@#1\endcsname
\fi
#2}}
\providecommand{\BIBdecl}{\relax}
\BIBdecl

\bibitem{Mitola}
J.~Mitola~III, ``Cognitive radio an integrated agent architecture for software
  defined radio,'' Ph.D. dissertation, KTH Royal Institute of Technology, May
  2000.

\bibitem{xG}
I.~F. Akyildiz, W.-Y. Lee, M.~C. Vuran, and S.~Mohanty, ``Ne{X}t
  generation/dynamic spectrum access/cognitive radio wireless networks: A
  survey,'' \emph{Computer Networks}, vol.~50, pp. 2127--2159, Sep. 2006.

\bibitem{relay}
E.~C. van~der Meulen, ``Three-terminal communication channels,'' \emph{Advances
  in Applied Probability}, vol.~3, no.~1, pp. 120--154, Spring, 1971.

\bibitem{A_relay}
W.~Jaafar, W.~Ajib, and D.~Haccoun, ``Opportunistic adaptive relaying in
  cognitive radio networks,'' in \emph{IEEE International Conference on
  Communications (ICC)}, Ottawa, ON, Jun. 2012, pp. 1811--1815.

\bibitem{B_relays}
Y.~Zou, J.~Zhu, B.~Zheng, and Y.-D. Yao, ``An adaptive cooperation diversity
  scheme with best-relay selection in cognitive radio networks,'' \emph{IEEE
  Transactions on Signal Processing}, vol.~58, no.~10, pp. 5438--5445, Oct.
  2010.

\bibitem{AF}
Y.~Han, A.~Pandharipande, and S.~H. Ting, ``Cooperative spectrum sharing via
  controlled amplify-and-forward relaying,'' in \emph{IEEE 19th International
  Symposium on Personal, Indoor and Mobile Radio Communications (PIMRC)},
  Cannes, France, Sep. 2008.

\bibitem{ST_DF}
------, ``Cooperative decode-and-forward relaying for secondary spectrum
  access,'' \emph{IEEE Transactions on Wireless Communications}, vol.~8,
  no.~10, pp. 4945--4950, Oct. 2009.

\bibitem{BC}
D.~Nguyen, T.~Tran, T.~Nguyen, and B.~Bose, ``Wireless broadcast using network
  coding,'' \emph{IEEE Transactions on Vehicular Technology}, vol.~58, no.~2,
  pp. 914--925, Feb. 2009.

\bibitem{RelayMC}
P.~Fan, C.~Zhi, C.~Wei, and K.~B. Letaief, ``Reliable relay assisted wireless
  multicast using network coding,'' \emph{IEEE Journal on Selected Areas in
  Communications}, vol.~27, no.~5, pp. 749--762, Jun. 2009.

\bibitem{BK98}
Y.~Birk and T.~Kol, ``Informed-source coding-on-demand ({ISCOD}) over broadcast
  channels,'' in \emph{Seventeenth Annual Joint Conference of the IEEE Computer
  and Communications Societies (INFOCOM)}, vol.~3, San Francisco, CA, Mar.-Apr.
  1998, pp. 1257--1264.

\bibitem{BK06}
------, ``Coding on demand by an informed source ({ISCOD}) for efficient
  broadcast of different supplemental data to caching clients,'' \emph{IEEE
  Transactions on Information Theory}, vol.~52, no.~6, pp. 2825--2830, Jun.
  2006.

\bibitem{LXR_VTC14}
N.~Li, M.~Xiao, and L.~K. Rasmussen, ``Cooperation-based network coding in
  cognitive radio networks,'' in \emph{2014 IEEE 80th Vehicular Technology
  Conference (VTC Fall)}, Vancouver, Canada, Sep. 2014.

\bibitem{OFDM_packet}
Y.~J. Zhang and K.~B. Letaief, ``Adaptive resource allocation and scheduling
  for multiuser packet-based {OFDM} networks,'' in \emph{IEEE International
  Conference on Communications (ICC)}, vol.~5, Paris, France, Jun. 2004, pp.
  2949--2953.

\bibitem{Larsson08wcnc}
P.~Larsson, ``Multicast multiuser {ARQ},'' in \emph{IEEE Wireless Commun. \&
  Networking Conf.}, Las Vegas, USA, Mar. 2008, pp. 1985--1990.

\bibitem{B66}
J.~J. Bartko, ``Approximating the negative binomial,'' \emph{Technometrics},
  vol.~8, no.~2, pp. 345--350, May 1966.

\bibitem{JB14}
\BIBentryALTinterwordspacing
J.~Burkardt. (2014, October) The truncated normal distribution. Department of
  Scientific Computing, Florida State University. [Online]. Available:
  \url{http://people.sc.fsu.edu/~jburkardt/presentations/truncated normal.pdf}
\BIBentrySTDinterwordspacing

\bibitem{CUD94}
N.~Johnson, S.~Kotz, and N.~Balakrishnan, \emph{Continuous Univariate
  Distributions}, second edition~ed., ser. Wiley Series in Probability and
  Mathematical Statistics.\hskip 1em plus 0.5em minus 0.4em\relax John Wiley \&
  Sons, Inc., 1994.

\bibitem{PM_PB95}
P.~Billingsley, \emph{Probability and Measure}, third edition~ed., ser. Wiley
  Series in Probability and Mathematical Statistics.\hskip 1em plus 0.5em minus
  0.4em\relax John Wiley \& Sons, Inc., 1995.

\bibitem{NW06}
N.~A. Weiss, P.~T. Holmes, and M.~Hardy, \emph{A Course in Probability}.\hskip
  1em plus 0.5em minus 0.4em\relax Pearson Addison Wesley, 2006.

\bibitem{C61}
C.~E. Clark, ``The greatest of a finite set of random variables,''
  \emph{Operations Research}, pp. 145--162, Mar.-Apr. 1961.

\bibitem{NK08}
S.~Nadarajah and S.~Kotz, ``Exact distribution of the max/min of two {G}aussian
  random variables,'' \emph{IEEE Transactions on Very Large Scale Integration
  (VLSI) Systems}, vol.~16, no.~2, pp. 210--212, Feb. 2008.

\end{thebibliography}

\end{document}